\documentclass[preprint, sort&compress]{elsarticle}
\usepackage{graphicx,graphics}
\usepackage{epstopdf}
\usepackage{dcolumn}
\usepackage{amsmath,amssymb,amsfonts}
\usepackage{latexsym,verbatim}
\usepackage{bm}
\usepackage{textcomp} 
\usepackage{mathtools}
\usepackage{braket}
\usepackage{soul}
\usepackage{bbold}
\usepackage{color}
\usepackage{ulem}
\usepackage[breaklinks=false,colorlinks,citecolor=blue,linkcolor=blue,urlcolor=blue]{hyperref}
\usepackage[abs]{overpic}
\usepackage{tikz, bm}
\usepackage[compat=1.1.0]{tikz-feynman}
\DeclareRobustCommand{\rchi}{{\mathpalette\irchi\relax}}
\newcommand{\irchi}[2]{\raisebox{\depth}{$#1\chi$}} 
\begin{document}
\title{Gauge invariance and Ward identities in nonlinear response theory}
\author[SU]{Habib Rostami}
\ead{habib.rostami@su.se}
\author[RU]{Mikhail I. Katsnelson}
\author[MIS]{Giovanni Vignale}
\author[P1,IT,P2]{Marco Polini}
\address[SU]{Nordita, KTH Royal Institute of Technology and Stockholm University, Stockholm SE-106 91, Sweden}
\address[RU]{Institute for Molecules and Materials, Radboud University, Heyendaalseweg 135, 6525 AJ, Nijmegen, The Netherlands}
\address[MIS]{Department of Physics and Astronomy, University of Missouri, Columbia, Missouri 65211,~USA}
\address[P1]{Dipartimento di Fisica dell'Universit\`a di Pisa, Largo Bruno Pontecorvo 3, I-56127 Pisa,~Italy}
\address[IT]{Istituto Italiano di Tecnologia, Graphene Labs, Via Morego 30, I-16163 Genova, Italy}
\address[P2]{School of Physics \& Astronomy, University of Manchester, Oxford Road, Manchester M13 9PL,~United Kingdom}

\date{\today}

\date{\today}

\begin{abstract}
We present a formal analysis of {\it nonlinear} response functions in terms of correlation functions in real- and imaginary-time domains.  In particular, we show that causal nonlinear response functions, expressed in terms of nested commutators in real time, can be obtained from the analytic continuation of time-ordered response functions, which are more easily amenable to diagrammatic calculation. This generalizes the well-known result of linear response theory.  We then use gauge invariance arguments to derive exact relations between second-order response functions in density and current channels. These identities, which are non-perturbative in the strength of inter-particle   interactions, allow us to establish exact connections between nonlinear optics calculations done in different electromagnetic gauges.
\end{abstract}


\maketitle

\section{Introduction}
\label{sec:Introduction}
Nonlinear optical phenomena~\cite{Landau08,Shen,Butcher_and_Cotter,Boyd}, arise from the response of an electronic system to intense electromagnetic fields. Going beyond linear response, many interesting optical effects can occur including harmonic generation, wave mixing, and saturable absorption~\cite{Butcher_and_Cotter}. As in the case of linear response theory, calculations of nonlinear response functions to an electromagnetic field can be carried out in different electromagnetic gauges such as the scalar potential gauge (in which the scalar potential $\Phi$ is finite, while the vector potential is ${\bm {\mathcal A}}=0$) and the vector potential gauge $(\Phi=0,{\bm {\mathcal A}} \neq {\bm 0})$. 
Nonlinear response functions are usually expressed in terms of complicated and lengthy relations, which are cumbersome to evaluate~\cite{Shen,Butcher_and_Cotter,Boyd}.
Therefore, one often needs to introduce some approximations in order to handle technical difficulties. For example, in the so-called ``optical" limit, the photon momentum is neglected and one needs to study only the {\it local} response to a homogenous time-dependent electric field. This treatment is often called electric-dipole approximation~\cite{Shen,Butcher_and_Cotter}.  In this approximation, the field-particle interaction can be written in two major gauges: ${\cal H}_{\Phi}={\cal H}({\bm p})-e{\bm E}(t)\cdot{\bm r}$ or  ${\cal H}_{\bm {\mathcal A}}={\cal H}({\bm p}+e{\bm {\mathcal A}}(t))$, where ${\cal H}$ is the Hamiltonian in the absence of radiation, and ${\bm r}$ and ${\bm p} \equiv -i\hbar {\bm \nabla}_{\bm r}$ stand for the position and momentum operators, respectively. The equivalence of these two gauges for an arbitrary physical observable in the electric-dipole approximation is well discussed in the literature~\cite{Genkin_Mednis_1968,Aversa_prb_1995,naguleswaran_jphysb_1996,bandrauk_jpb_2013,Ventura_arXiv_2017}. 
\par
For optical properties of insulators, we mainly need to consider inter-band transitions while the intra-band contribution is more relevant in metals and semimetals. To calculate the inter-band optical response, it is convenient to utilize the vector-potential gauge~\cite{anti_arXiv_2016} in which one can simply drop the photon momentum from the very beginning of the calculation. Apart from pure intra- and inter-band transitions, an extra class of transitions emerges {\it only} in the analysis of nonlinear response functions containing $n$-point correlation functions with $n\ge3$.  This new contribution originates from mixed transitions which contain both intra- and inter-band processes~\cite{mikhailov_prb_2016,rostami_prb_2016}. 
Employing a generic density gauge, $\Phi({\bm r},t)$, is more suitable for evaluating the contributions due to all intra-band, inter-band and mixed transitions in the optical limit~\cite{mikhailov_prb_2016,rostami_prb_2016,rostami_prb_2017}. This is because only a single non-interacting Feynman diagram is required to be evaluated in the density gauge at any order of perturbation theory, while in the vector potential gauge there are two, four, and eight diagrams that need to be evaluated for the linear, second-order and third-order response functions, respectively~\cite{Moore_prb_2019}. 
\par
In nonlinear response theory, the gauge choice is crucial when going beyond the electric-dipole approximation. If one keeps the photon momentum up to linear order in the nonlinear conductivity, it is possible to capture nonlocal effects in the electric-quadrupole and magnetic-dipole approximations~\cite{rostami_prb_2017,cheng_sr_2017}. These new contributions are very important for the second-order nonlinear response of inversion symmetric systems~\cite{Butcher_and_Cotter,apell_physicascripta_1983,chernyak_jchemphys_2015}. 
Note that in the scalar potential gauge, one loses the transverse contribution to the current, which originates from the magnetic-dipole moment~\cite{cheng_sr_2017}. In fact, as discussed in this Article, in order to calculate the nonlocal current it is preferable to perform the calculation in the vector potential gauge by considering a spatially inhomogeneous vector field, ${\bm {\mathcal A}}({\bm r},t)$. 
\par
As just reviewed, most of the available literature on the issue of gauge invariance in nonlinear response theory is devoted to exploring the gauge choice (the two specific aforementioned gauges in the electric-dipole approximation) in non-interacting atoms~\cite{Lamb_pra_1987} and crystalline solids~\cite{Aversa_prb_1995}. To the best of our knowledge, a generic analysis of gauge invariance for a spatially inhomogeneous external field in nonlinear response theory as applied to {\it interacting} electron systems is still unexplored and it is one of the main motivations of this work. Our analysis of nonlinear response functions covers all local (electric-dipole) and nonlocal (other electric and magnetic-multipoles) effects and can be seen as a generalization of the gauge invariance analysis in the context of linear response theory, which is textbook material~\cite{Schrieffer,Pines_and_Nozieres,Giuliani_and_Vignale}.
Nonlocal response functions are also of interest when the momentum exchanged between photons (external field) and charged particles cannot be ignored. This is the case of the photon-drag effect~\cite{Grinberg_prb_1988,Glazov_prep_2014,Lundeberg_science_2017} and of other effects where the photon momentum needs to be taken into account. For example, our analysis may be relevant to analyze experimental results of spectroscopy based on free-electron lasers~\cite{Hopf_OC_1976,McNeil_2010}.
\par

Gauge invariance, i.e.~the independence of physical observables on the electromagnetic gauge one chooses for calculations, imposes severe constraints on the theory of linear response~\cite{Schrieffer,Pines_and_Nozieres,Giuliani_and_Vignale}. Indeed, 1) the fact that a static and purely longitudinal vector potential cannot produce any physical current implies the vanishing of the {\it longitudinal} current response at any finite wave vector $q$; 2) similarly, the fact that a static and quasi-homogeneous vector potential cannot produce any physical current implies the vanishing of the longitudinal {\it and} transverse current-current response functions for $q \to 0$. More generally, gauge invariance imposes precise relationships between linear response functions to scalar and vector potentials.
\par
In this Article, we first present a formal analysis of the nonlinear response theory in the real and imaginary time domains. We show that causal nonlinear response functions, expressed in terms of nested commutators in real time, can be obtained from the analytic continuation of time-ordered response functions, which are more easily amenable to diagrammatic calculation. This generalizes the well-known result of linear response theory.

We then provide a theoretical study of the impact of gauge invariance on the second-order response functions. We finally report Ward identities which must be fulfilled in order to ensure gauge invariance at any order of perturbation theory.  
In Section~\ref{sec:nonlinear_response}, we report an explicit definition of nonlinear response functions.
In Section~\ref{sec:Lehmann}, we discuss the spectral  representation of second-order response functions in real- and imaginary-time domains. 
In Section~\ref{sec:light_matter}, we consider a specific kind of external field, i.e.~electromagnetic radiation treated classically. 
In Sections~\ref{sec:linear} and~\ref{sec:second}, we explicitly discuss gauge invariance for linear and second-order response functions, respectively.  
Finally, in Section~\ref{sec:ward}, we present a set of Ward identities for nonlinear response functions.

\section{Nonlinear response  theory in real time}\label{sec:nonlinear_response}
\begin{figure}[t]
\centering
\begin{overpic}[width=0.9\linewidth]{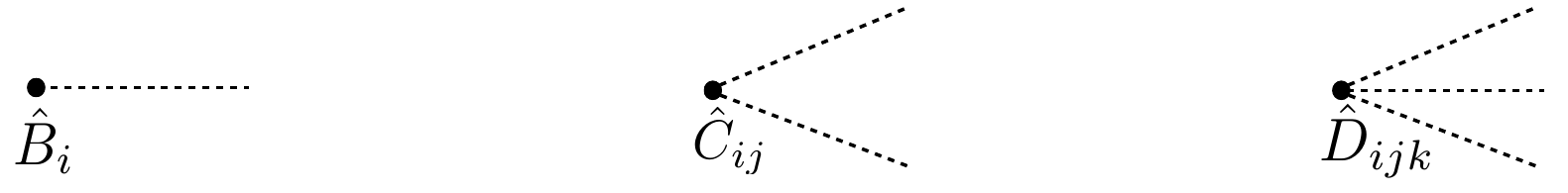}\put(10,70){}\end{overpic}
\caption{Diagrammatic representation of the linear and nonlinear couplings as vertices (solid dots) coupled to external fields (dashed lines). In the case of electromagnetic perturbations, we call the $  B$, $  C$, and $  D$ vertices as single-photon, two-photon and three-phonon vertices, respectively.}
\label{fig:vertex}
\end{figure}

We consider a many-body system in thermal equilibrium which is described by a Hamiltonian denoted by $\hat{\cal H}$. We then turn on an external field, which can be modelled by a field-particle interaction term 
denoted by $\hat{V}$. We use the $\hat{S}$-matrix approach~\cite{Mahan} to consider the effect of the interaction in a perturbative way. In this approach, one has an adiabatic time evolution of the wave function from an unperturbed state at time $t=-\infty$, i.e.~$|\psi_0\rangle$, to the perturbed state at time $t$. Therefore, we have $|\psi(t)\rangle= \hat{S}(t,-\infty) |\psi_0\rangle$ where the $\hat{S}$-matrix is given by (setting $\hbar=1$ for the sake of simplicity)
\begin{equation}
\hat S(t,-\infty)= {\cal T} \exp \left (-i\int^t_{-\infty}  \hat V(t') dt' \right )~.
\end{equation}
Here, ${\cal T}$ stands for the time-ordering operation and $\hat{V}(t)$ is the perturbative part of the Hamiltonian in the interaction representation, i.e.~$\hat V(t) =e^{it \hat{\cal H}} \hat V e^{-it \hat{\cal H}}$. In a similar way, any observable evolves in the interaction picture as 
\begin{equation}\label{eq:interaction_picture}
\hat {A}(t) = e^{it \hat{\cal H}} \hat {A} e^{-it \hat{\cal H}}~. 
\end{equation}
Here, $\hat{A}$ denotes the second-quantized representation of the ${A}$ operator.  The expectation value of $\hat {A}$ at time $t$ is given by
\begin{equation}\label{eq:O_ave_1}
A(t) =\langle \psi(t) | \hat {A}(t)|\psi(t)\rangle=\langle \psi_0 | \hat S^\dagger(t,-\infty) \hat {A}(t) \hat S(t,-\infty)|\psi_0\rangle~.
\end{equation}
Thanks to the Dyson expansion, we have (see~\ref{app:Dyson_series})
\begin{align}\label{eq:perturbation_series_A}
{A}(t) &= \left\langle \hat {A}(t)\right \rangle + i\int^{\infty}_{-\infty}d\tau_1\Theta(\tau_1) \left\langle\left[\hat V(t-\tau_1), \hat {A}(t)\right]\right \rangle
\nonumber\\ 
&+
i^2 \int^{\infty}_{-\infty} d\tau_1\int^{\infty}_{-\infty}d\tau_2~ \Theta(\tau_2-\tau_1) \Theta(\tau_1) \left\langle\left[\hat V(t-\tau_2),\left[\hat V(t-\tau_1), \hat {A}(t)\right]\right]\right \rangle
\nonumber\\ 
&+
i^3\int^{\infty}_{-\infty}d\tau_1\int^{\infty}_{-\infty}d\tau_2 \int^{\infty}_{-\infty} d\tau_3 ~ \Theta(\tau_3-\tau_2)\Theta(\tau_2-\tau_1)  \Theta(\tau_1)  
\nonumber\\ 
&\times
\left\langle \left[\hat V(t-\tau_3),\left[\hat V(t-\tau_2),\left[\hat V(t-\tau_1), \hat {A}(t)\right]\right]\right]\right \rangle
+\dots~,
\end{align}
where $\Theta(\tau)$ is the Heaviside function. The field-particle interaction can be formally written as follows  
\begin{equation}\label{eq:V_BF}
\hat V(t) = \sum_{i} \hat B_i (t) F_ i (t) + \sum_{ij} \hat C_{ij} (t) F_ i (t) F_ j (t) +\sum_{ijk} \hat D_{ijk} (t) F_ i (t) F_ j (t) F_ k (t) + \dots~,
\end{equation}
where $\hat B_i (t)$ indicates linear coupling while $ \hat C_{ij} (t)$ and $ \hat D_{ijk} (t)$ stand for the nonlinear couplings. 
We can visualize these linear and nonlinear coupling terms as vertices in the diagrams shown in Fig.~\ref{fig:vertex}. 
For the case of light-matter interactions, we term the $B$, $C$, and $D$ vertices as single-photon, two-photon, and three-phonon vertices, respectively.  
Here, we first assume a linear coupling via $\hat B_i$ to the external fields $F_i$ and later generalize our formal theory to the nonlinear couplings $\hat C_{ij} (t)$ and $\hat D_{ijk} (t)$. 
Note that vertex couplings like $\hat C_{ij} (t)$ and $\hat D_{ijk} (t)$ only emerge in the non-relativistic limit of a low-energy theory because of the nonlinear energy dispersion of the band Hamiltonian ($\epsilon_{\bm k} \sim \sum_{n\ge 0}\epsilon_n k^n$) and they do not exist in quantum electrodynamics~\cite{Peskin_and_Schroeder}.  
After neglecting nonlinear coupling terms and plugging Eq.~(\ref{eq:V_BF}) in Eq.~(\ref{eq:perturbation_series_A}), we obtain 
\begin{align}\label{eq:A_ave1}
 A(t) &= \left\langle \hat A(t)\right \rangle+ \sum_{i} \int^{\infty}_{-\infty}d\tau_1~\rchi_{A B_i} (t; \tau_1)  F_i (t-\tau_1)
\nonumber \\
&+ \sum_{ij} \int^{\infty}_{-\infty} d\tau_1\int^{\infty}_{-\infty}d\tau_2~\rchi_{A B_i B_j} (t;\tau_1,\tau_2)    F_i (t-\tau_1) F_j (t-\tau_2)
\nonumber\\ 
&+
\sum_{ijk} \int^{\infty}_{-\infty}d\tau_1\int^{\infty}_{-\infty}d\tau_2 \int^{\infty}_{-\infty} d\tau_3 ~ \rchi_{A B_i B_j B_k}(t;\tau_1,\tau_2,\tau_3) 
\nonumber \\ 
&\times
F_i (t-\tau_1)  F_j (t-\tau_2)  F_k (t-\tau_3) 
+\dots~.
\end{align}
The previous expression leads us to define the linear {\it retarded} response functions~\cite{Giuliani_and_Vignale}
\begin{equation}\label{eq:chi1_t}
\rchi_{A B_i} (t; \tau_1) = i  \Theta(\tau_1)  \left \langle\left[\hat B_i(t-\tau_1), \hat A(t)\right] \right \rangle~,
\end{equation}
the second-order retarded response functions
\begin{equation}\label{eq:chi2_t}
\rchi_{A B_i B_j} (t;\tau_1,\tau_2) = i^2 \sum'_{\cal P}  \Theta(\tau_2-\tau_1) \Theta(\tau_1) \left\langle\left[\hat B_j (t-\tau_2),\left[\hat B_i (t-\tau_1), \hat A(t)\right]\right]\right \rangle~,
\end{equation}
and the third-order ones 
\begin{align}\label{eq:chi3_t}
\rchi_{A B_i B_j B_k}(t;\tau_1,\tau_2,\tau_3) &=  i^3 \sum'_{\cal P} \Theta(\tau_3-\tau_2)\Theta(\tau_2-\tau_1)  \Theta(\tau_1) 
\nonumber \\ &\times
 \left\langle \left[\hat B_k(t-\tau_3),\left[\hat B_j(t-\tau_2),\left[\hat B_i(t-\tau_1), \hat A(t)\right]\right]\right]\right \rangle~.
\end{align}
Of course, one can simply extend these definitions to higher-order response functions. 
The symbol $\sum'_{\cal P}$ in Eqs.~(\ref{eq:chi2_t}) and (\ref{eq:chi3_t}) is there to ensure a permutation symmetry. Since the quantity $\Pi^m_{j=1}F_{j}(t-\tau_j)$ in Eq.~(\ref{eq:A_ave1}) is symmetric with respect to the permutation between each pair of dummy variables, i.e.~$i\equiv (F_i,\tau_i)$ and $j\equiv (F_j,\tau_j)$, we expect a permutation symmetry for the $m$-th order response function. For instance, following Ref.~\cite{Butcher_and_Cotter}, one can decompose the $m$-th order response function in the sum of symmetric and anti-symmetric contributions, i.e.~${\bm \rchi}={\bm \rchi}^{\rm s}+{\bm \rchi}^{\rm a}$ with
\begin{align}
\rchi^{\rm s}_{A B_1 B_2 \dots B_m} (t;\tau_1,\tau_2,\dots,\tau_m)&=\frac{1}{2} 
\Big \{ 
\rchi_{A B_1 B_2 \dots B_m} (t;\tau_1,\tau_2,\dots,\tau_m)
\nonumber \\ &+
\rchi_{A B_2 B_1 \dots B_m} (t;\tau_2,\tau_1,\dots,\tau_m)
\Big \}
\end{align}
and
\begin{align}
\rchi^{\rm a}_{A B_1 B_2 \dots B_m} (t;\tau_1,\tau_2,\dots,\tau_m)&=
\frac{1}{2} 
\Big \{ 
\rchi_{A B_1 B_2 \dots B_m} (t;\tau_1,\tau_2,\dots,\tau_m)
\nonumber \\ &-
\rchi_{A B_2 B_1 \dots B_m} (t;\tau_2,\tau_1,\dots,\tau_m)
\Big \}~.
\end{align}
It is obvious that ${\bm \rchi}^{\rm s}$ (${\bm \rchi}^{\rm a}$) is symmetric (anti-symmetric) with respect to the permutation between ($B_1,\tau_1$) and ($B_2,\tau_2$). 
Since the term $\Pi^m_{j=1}F_{j}(t-\tau_j)$ is symmetric for all orders of permutation, the anti-symmetric part plays no role in determining the expectation value of $\hat A$. 
The only non-vanishing contribution to the latter originates from completely symmetric response functions. This is the reason why in the definition of the $m$-th order nonlinear response function we have introduced the sum $\sum'_{\cal P}=\frac{1}{m!} \sum_{\cal P}$, where $\sum_{\cal P}$ stands for the sum over all permutations among the dummy variables ($B_i,\tau_i$). This kind of symmetry is usually called {\it intrinsic permutation symmetry}~\cite{Butcher_and_Cotter}. 

By using the cyclic properties of the trace, we can then eliminate $``t"$ in Eqs.~(\ref{eq:chi1_t}), (\ref{eq:chi2_t}), and (\ref{eq:chi3_t}):
\begin{align}
\rchi_{A B_i} (\tau_1) &= i  \Theta(\tau_1)  \left \langle\left[\hat B_i(-\tau_1), \hat A(0)\right] \right \rangle~, \label{eq:chi1}
\\
\rchi_{A B_i B_j} (\tau_1,\tau_2) &= i^2  \sum'_{\cal P}\Theta(\tau_2-\tau_1) \Theta(\tau_1) 
\left\langle\left[\hat B_j (-\tau_2),\left[\hat B_i (-\tau_1), \hat A(0)\right]\right]\right \rangle~,\label{eq:chi2}
\\
\rchi_{A B_i B_j B_k}(\tau_1,\tau_2,\tau_3)  &= i^3  \sum'_{\cal P} \Theta(\tau_3-\tau_2)\Theta(\tau_2-\tau_1)  \Theta(\tau_1)  
\nonumber \\ &\times
\left\langle \left[\hat B_k(-\tau_3),\left[\hat B_j(-\tau_2),\left[\hat B_i(-\tau_1), \hat A(0)\right]\right]\right]\right \rangle~.\label{eq:chi3}
\end{align}
The invariance under time translations of the previous response functions is evident. They just depend on the time differences $\tau_i=t-t_i$. This can be powerfully used by Fourier transforming to the frequency domain:
\begin{align}
\rchi_{A B_i} (\omega_1) &=  \int^{\infty}_{-\infty}d\tau ~\rchi_{A B_i} (\tau ) e^{i(\omega_1+i\eta_1) \tau}~,\label{eq:chi1_FQ}
\\
\rchi_{A B_i B_j} (\omega_1,\omega_2) &= \int^{\infty}_{-\infty} d\tau_1\int^{\infty}_{-\infty}d\tau_2~\rchi_{A B_i B_j} (\tau_1,\tau_2)  e^{i (\omega_\Sigma +i\sum_i\eta_i)\tau_i}~,\label{eq:chi2_FQ}
\\
\rchi_{A B_i B_j B_k}(\omega_1,\omega_2,\omega_3) &= \int^{\infty}_{-\infty} d\tau_1\int^{\infty}_{-\infty}d\tau_2 \int^{\infty}_{-\infty}d\tau_3~\rchi_{A B_i B_j B_k}(\tau_1,\tau_2,\tau_3)  
\nonumber\\&\times e^{i (\omega_\Sigma+i\sum_i\eta_i) \tau_i}~.
\label{eq:chi3_FQ}
\end{align}
Here, $\omega_\Sigma=\sum_i \omega_i$. 
As usual~\cite{Giuliani_and_Vignale}, we have assumed that all frequencies contain an infinitesimal positive imaginary part $\eta_i \to 0^+$. This stems from the need to fit periodic perturbations $F_{i}(t)=F_{i}(\omega)e^{-i\omega t}e^{-\eta t}+{\rm c.c.}$ into the response theory formalism, making sure that these vanish in the remote past.
In the frequency domain, the  intrinsic permutation symmetry is among the ($B_i,\omega_i$) pairs.
The thermal average in Eqs.~(\ref{eq:chi1}), (\ref{eq:chi2}), and (\ref{eq:chi3}) is defined as usual~\cite{Giuliani_and_Vignale}, i.e.~
$\langle \hat {\cal O}  \rangle = {\rm Tr}  [ \hat \rho \hat{\cal O}]=\sum_{\lambda} P_\lambda \langle \lambda | \hat {\cal O} |\lambda \rangle$
where $\hat \rho$ is the density matrix, $P_\lambda= \langle \lambda| \hat \rho |\lambda \rangle =  {e^{-\beta E_\lambda}}/{\cal Z}$, ${\cal Z}=\sum_\lambda e^{-\beta E_\lambda}$ is the partition function, and $\beta=1/(k_{\rm B} T)$ with $T$ the temperature. Notice that $|\lambda \rangle$ and $E_\lambda$ are the {\it exact} eigenstates and eigenvalues of the many-body Hamiltonian $\hat{\cal H}$ in the absence of radiation. The same assumptions that are made on $P_\lambda$ in the linear-response Kubo formalism (see footnote 11 in Chapter 3 of Ref.~\cite{Giuliani_and_Vignale}) are assumed to hold true in this nonlinear case too.

\begin{figure}[t]
\centering
\begin{overpic}[width=0.45\linewidth]{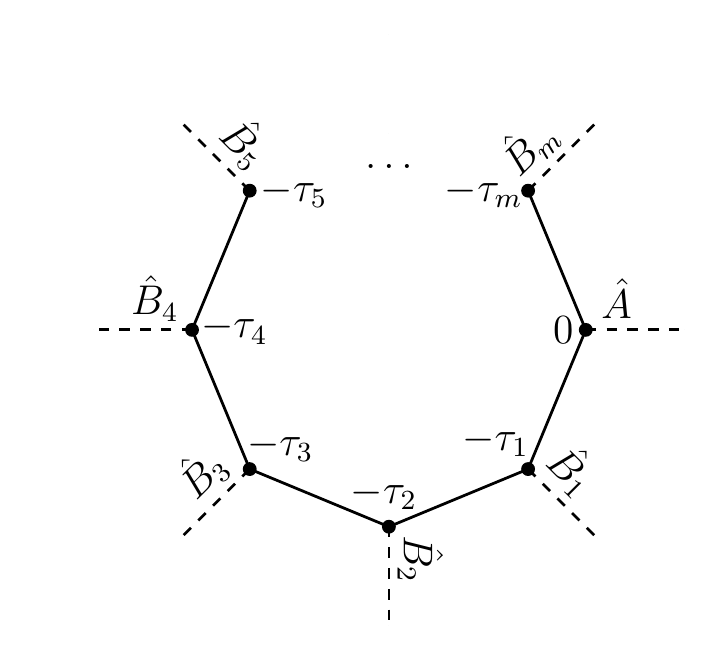}\put(5,100){\large (a)}\end{overpic}
\begin{overpic}[width=0.45\linewidth]{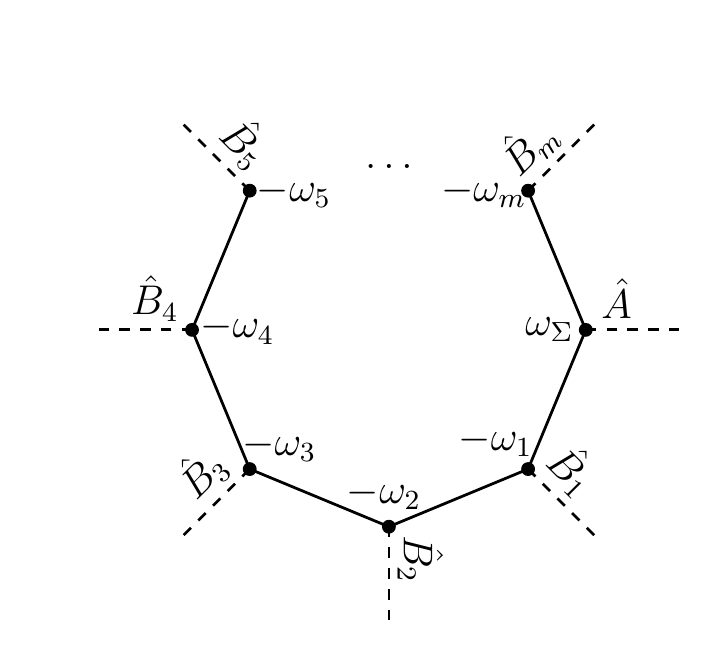}\put(5,100){\large (b)}\end{overpic}
\caption{Feynman diagram for the $m$-th order response function in the time, panel (a), and frequency, panel (b), domains.
These diagrams just stand for one of the $m!$ possible permutations among the dummy variables. Notice that $\omega_\Sigma \equiv \sum^m_{i=1} \omega_i$~.}
\label{fig:nth_diagrams}
\end{figure}

With the definitions given in Eqs.~(\ref{eq:chi1}), (\ref{eq:chi2}), and (\ref{eq:chi3}), we can take advantage of diagrammatic techniques to evaluate nonlinear response functions. In Fig.~\ref{fig:nth_diagrams}, we give a diagrammatic representation of the $m$-th order response function in both time and frequency domains. 
The generalization of the linear-coupling theory to the nonlinear couplings is straightforward as it is simply based on the inclusion of higher-order photon vertices in the Feynman diagrams.  
\section{Spectral representation and analytic continuation}\label{sec:Lehmann}

In an interacting many-body system, it is not easy to handle all the diagrams for linear and nonlinear response functions. It is often convenient to do the calculations with time-ordered response functions in imaginary time and then come back to real time by an analytic continuation~\cite{Mahan}.  The imaginary-time representation of response functions, also known as ``Matsubara representation",  is also extremely useful for the finite temperature analysis. 
In order to derive the connection between causal (real-time) response function and time-ordered response functions in imaginary time, we proceed in three steps.  First we derive the so-called spectral representation (also known as ``Lehmann representation" or ``exact  eigenstates representation") for second-order response functions.  Next, making use of this representation, it is shown that the second-order causal response functions in real time are connected to time-ordered second-order response functions in imaginary time by the well-known analytic continuation procedure~\cite{Mahan}. Finally, reasoning by induction, we prove that the analytic continuation procedure works for response functions of arbitrary order.

\subsection{Real time (frequency) spectral representation: second-order response}
According to Eq.~(\ref{eq:chi2}), the second-order retarded response function in the real-time representation reads as follows: 
\begin{align}
\rchi_{A B_1 B_2} (\tau_1,\tau_2) 
&= i^2  \sum'_{\cal P}\Theta(\tau_2-\tau_1) \Theta(\tau_1) 
\left\langle\left[\hat B_2 (-\tau_2),\left[\hat B_1 (-\tau_1), \hat A(0)\right]\right]\right \rangle
\nonumber \\
&= i^2 \sum'_{\cal P}\Theta(\tau_2-\tau_1) \Theta(\tau_1) I_{AB_1B_2}(\tau_1,\tau_2)~,
\end{align}
where
\begin{align}
I_{AB_1B_2}(\tau_1,\tau_2)&= 
 \left\langle \hat B_2 (-\tau_2) \hat B_1 (-\tau_1)  \hat A(0)\right \rangle 
- \left\langle\hat B_1 (-\tau_1)  \hat A(0) \hat B_2 (-\tau_2)\right \rangle
\nonumber\\
&-\left\langle\hat B_2 (-\tau_2)   \hat A(0) \hat B_1 (-\tau_1)\right \rangle
+\left\langle \hat A(0) \hat B_1 (-\tau_1) \hat B_2 (-\tau_2)\right \rangle~.
\end{align}
Performing the thermodynamic average, we obtain 
\begin{align}
I_{AB_1B_2}(\tau_1,\tau_2)
&=\sum_{ \lambda_1  \lambda_2  \lambda_3} P_{ \lambda_1}
 \Bigg \{ 
\langle  \lambda_1 | \hat B_2(-\tau_2) | \lambda_2 \rangle \langle  \lambda_2| \hat B_1(-\tau_1) | \lambda_3 \rangle \langle  \lambda_3 | \hat A(0) |\lambda_1 \rangle
\nonumber\\
&-\langle  \lambda_1 | \hat B_1(-\tau_1)| \lambda_2 \rangle \langle \lambda_2 | \hat A(0) | \lambda_3 \rangle \langle  \lambda_3 | \hat B_2(-\tau_2)   |\lambda_1 \rangle
\nonumber \\
&-\langle  \lambda_1 | \hat B_2(-\tau_2)  |\lambda_2 \rangle \langle \lambda_2 |\hat A(0) | \lambda_3 \rangle \langle  \lambda_3 |\hat B_1(-\tau_1) |\lambda_1 \rangle
\nonumber\\
&+\langle   \lambda_1 |  \hat A(0) | \lambda_2 \rangle \langle \lambda_2 | \hat B_1(-\tau_1) | \lambda_3 \rangle \langle  \lambda_3 | \hat B_2(-\tau_2)|\lambda_1 \rangle
 \Bigg \}~.
\end{align}
We exchange $\lambda_1$, $\lambda_2$  and $\lambda_3$ dummy labels in the second term of the above relation in order to the collect it with the first term. We use a similar trick and
collect the last two terms together. Consequently, we arrive at 
\begin{align}
I_{AB_1B_2}(\tau_1,\tau_2) &=
    \sum_{\lambda_1  \lambda_2  \lambda_3} A_{\lambda_3 \lambda_1}
  \Bigg \{ P_{\lambda_1  \lambda_2} \langle \lambda_1 | \hat B_2(-\tau_2) |\lambda_2\rangle \langle \lambda_2| \hat B_1(-\tau_1) | \lambda_3 \rangle
 \nonumber \\ 
 &-
  P_{\lambda_2  \lambda_3} \langle \lambda_1 | \hat B_1(-\tau_1)  |\lambda_2\rangle \langle \lambda_2| \hat B_2(-\tau_2) | \lambda_3 \rangle
   \Bigg \}~.
\end{align}
Notice that $\langle  \lambda_3 | \hat A(0) |\lambda_1 \rangle = A_{\lambda_3 \lambda_1}$ and $P_{\lambda_1\lambda_2} =P_{\lambda_1}-P_{\lambda_2}$.
We extract the time-dependent part as follows
\begin{align}\label{eq:IAB1B2_time}
I_{AB_1B_2}(\tau_1,\tau_2) &= \sum_{\lambda_1  \lambda_2  \lambda_3} A_{\lambda_3\lambda_1}
  \Bigg \{ P_{\lambda_1\lambda_2}  B_{2,\lambda_1\lambda_2} B_{1,\lambda_2\lambda_3} e^{i\tau_2 E_{\lambda_2\lambda_1}} e^{i\tau_1 E_{\lambda_3\lambda_2}}
\nonumber \\
&-
  P_{\lambda_2\lambda_3}  B_{1,\lambda_1\lambda_2} B_{2,\lambda_2\lambda_3} e^{i\tau_1 E_{\lambda_2\lambda_1}} e^{i\tau_2 E_{\lambda_3\lambda_2}}              
   \Bigg \}~,
\end{align}
where $E_{\lambda_1\lambda_2} =E_{\lambda_1}-E_{\lambda_2}$. Using Eq.~(\ref{eq:chi2_FQ}), we obtain the frequency-domain representation of the second-order response function: 
\begin{align} \label{eq:chi2_omega}
\rchi_{A B_1 B_2} (\omega_1,\omega_2) 
= i^2\sum'_{\cal P} \int^{\infty}_{0} d\tau_1 \int^{\infty}_{\tau_1} d\tau_2 ~ I_{AB_1B_2}(\tau_1,\tau_2) e^{i\sum_i (\omega_i+i\eta_i)\tau_i}~.
\end{align}
Considering Eq.~(\ref{eq:IAB1B2_time}), we can easily perform the integrations in Eq.~(\ref{eq:chi2_omega}) and we reach   
\begin{align}
\rchi_{A B_1 B_2} (\omega_1,\omega_2) &= \sum'_{\cal P}  \sum_{\lambda_1  \lambda_2  \lambda_3} \frac{ A_{\lambda_3\lambda_1}}{\omega_1+\omega_2+E_{\lambda_3\lambda_1}+i [\eta_1+\eta_2]} \times 
\nonumber \\ &
\Bigg \{ 
B_{1,\lambda_1\lambda_2} B_{2,\lambda_2\lambda_3} \frac{P_{\lambda_3\lambda_2}}{ {\omega_2+E_{\lambda_3\lambda_2}+i\eta_2}} 
\nonumber \\
&-
B_{2,\lambda_1\lambda_2} B_{1,\lambda_2\lambda_3}  \frac{P_{\lambda_2\lambda_1}  }{ {\omega_2+E_{\lambda_2\lambda_1}+i\eta_2}}            
   \Bigg \}~.
\end{align}
Eventually, by considering the intrinsic permutation, we arrive at the following final expression for the retarded second-order response function:
\begin{align}\label{eq:chi2_ret}
\rchi_{A B_1 B_2} (\omega_1,\omega_2) &=
 \sum'_{\cal P}
   \sum_{\lambda_1  \lambda_2  \lambda_3} \frac{ A_{\lambda_{1}\lambda_{3}}B_{2,\lambda_{3}\lambda_{2}} B_{1,\lambda_{2}\lambda_{1}}}{\omega_1+\omega_2+E_{\lambda_{1}\lambda_{3}}+i [\eta_1+\eta_2]} 
  \nonumber \\ &\times
   \left [ 
    \frac{P_{\lambda_{1} \lambda_{2}}}{ {\omega_1+E_{\lambda_{1}\lambda_{2}}+i\eta_1}} - 
    \frac{P_{\lambda_{2}\lambda_{3}}  }{ {\omega_2+E_{\lambda_{2}\lambda_{3}}+i\eta_2}}  
   \right ]~.
\end{align}
We should set $\eta_1=\eta_2$ in order to turn on all the external fields at the same rate in the adiabatic switching-on procedure.

\subsection{Imaginary time (frequency) spectral representation: second-order response}

In the imaginary time, $\tau \equiv i t$, the time evolution of an operator obeys:
$\hat {\cal O}(\tau) = e^{\tau {\cal H}} \hat {\cal O} e^{-\tau {\cal H}}$~.
We should notice that only in this subsection $\tau$ represents an imaginary time. 
We generalize the periodic property of a bosonic correlation function, i.e. $f(\tau)=f(\tau+\beta)$,~\cite{Mahan} to a multi-variable correlation function as $f(\tau_1,\tau_2,\dots,\tau_m)$. In other words, for an arbitrary choice of $\tau_i$, we have 
$f(\tau_1,\tau_2,\dots,\tau_i,\dots,\tau_m) =f(\tau_1,\tau_2,\dots,\tau_i+\beta,\dots,\tau_m)$~.
Therefore, we can introduce the following Fourier and inverse-Fourier transformations:
\begin{align}\label{eq:Fourier_tau}
f(i\nu_{1},i\nu_{2},\dots,i\nu_{m}) &= \int^\beta_0 d\tau_1  \int^\beta_0 d\tau_2 \dots  \int^\beta_0 d\tau_m
 ~ f(\tau_1,\tau_2,\dots,\tau_m) e^{i\sum_i \nu_{i} \tau_i}~,
 \nonumber \\
f(\tau_1,\tau_2,\dots,\tau_m) &= \left (\frac{1}{\beta} \right)^m \sum_{\{\nu_i\}} f(i\nu_{1},i\nu_{2},\dots,i\nu_{m})  e^{-i\sum_i \nu_{i} \tau_i}~,
\end{align}
where $\nu_i= 0,  \pm 2\pi/\beta, \pm 4\pi/\beta \dots$ stands for the bosonic Matsubara frequency (energy).  
Now, we employ the Matsubara representation for the case of the second-order response in order to generalize the first-order Matsubara technique to higher-order response functions.  
The imaginary {\it time-ordered} second-order response function is {\it defined} by 
\begin{align}
\widetilde \rchi_{A B_1 B_2} (\tau_1,\tau_2) = \frac{(-1)^2}{2!}\left\langle { \cal T} \hat B_2 (-\tau_2) \hat B_1 (-\tau_1) \hat A(0) \right \rangle~.
\end{align}
After performing the time-ordering operation, we arrive at
\begin{align}
\widetilde \rchi_{A B_1 B_2} (\tau_1,\tau_2) &= \frac{1}{2} \Bigg \{
\Theta(\tau_1-\tau_2) \Theta(-\tau_1)  \left\langle \hat B_2 (-\tau_2) \hat B_1 (-\tau_1)  \hat A(0)\right \rangle 
\nonumber\\ &
+\Theta(\tau_2-\tau_1) \Theta(\tau_1)  \left\langle \hat A(0) \hat B_1 (-\tau_1) \hat B_2 (-\tau_2)\right \rangle
\nonumber\\ &
+\Theta(\tau_2-\tau_1) \Theta(-\tau_2)  \left\langle \hat B_1 (-\tau_1) \hat B_2 (-\tau_2)  \hat A(0)\right \rangle 
\nonumber\\ &
+\Theta(\tau_1-\tau_2) \Theta(\tau_2)  \left\langle \hat A(0) \hat B_2 (-\tau_2) \hat B_1 (-\tau_1)\right \rangle
\nonumber\\ &
+\Theta(\tau_2) \Theta(-\tau_1)   \left\langle\hat B_1 (-\tau_1)  \hat A(0) \hat B_2 (-\tau_2)\right \rangle
\nonumber\\ &
+\Theta(\tau_1) \Theta(-\tau_2)   \left\langle\hat B_2 (-\tau_2)   \hat A(0) \hat B_1 (-\tau_1)\right \rangle \Bigg \}~.
\end{align}
Since $0 \le \tau_i \le \beta$---see Eq.~(\ref{eq:Fourier_tau})--- we have $\Theta(-\tau_i)=0$ and  $\Theta(\tau_i)=1$ which implies
\begin{align}
\widetilde \rchi_{A B_1 B_2} (\tau_1,\tau_2) &=
\frac{1}{2} \Big \{ \Theta(\tau_2-\tau_1)   \left\langle \hat A(0) \hat B_1 (-\tau_1) \hat B_2 (-\tau_2)\right \rangle
\nonumber \\
&+\Theta(\tau_1-\tau_2)   \left\langle \hat A(0) \hat B_2 (-\tau_2) \hat B_1 (-\tau_1)\right \rangle \Big\}~.
\end{align}
The above equation can be rewritten in a compact form as follows
\begin{equation}
\widetilde \rchi_{A B_1 B_2} (\tau_1,\tau_2) = \sum'_{\cal P} \Theta(\tau_2-\tau_1)  \left\langle \hat A(0) \hat B_1 (-\tau_1) \hat B_2 (-\tau_2)\right \rangle~,
\end{equation}
where $\sum'_{\cal P}$ is the usual intrinsic permutation operation symbol. 
The thermodynamic average can be taken in the spectral representation:
\begin{align}
\widetilde \rchi_{A B_1 B_2} (\tau_1,\tau_2) 
&=\sum'_{\cal P} \sum_{\lambda_1 \lambda_2\lambda_3}  \Theta(\tau_2-\tau_1)  P_{\lambda_1}
 \langle \lambda_1| \hat A(0) |\lambda_2\rangle
  \langle \lambda_2| \hat B_1 (-\tau_1) |\lambda_3\rangle 
  \nonumber\\&
  \langle \lambda_3| \hat B_2 (-\tau_2) |\lambda_1\rangle~.
 \end{align}
According to the time-evolution convention and considering Eq.~(\ref{eq:Fourier_tau}) for the Fourier transformation definition in the imaginary-time domain, 
we Fourier transform $ \widetilde \rchi_{A B_1 B_2} (\tau_1,\tau_2)$ obtaining
\begin{align}
 \widetilde \rchi_{A B_1 B_2} (i\nu_1,i\nu_2)  &=  \sum'_{\cal P} \sum_{\lambda_1 \lambda_2\lambda_3} P_{\lambda_1}
  A_{\lambda_1\lambda_2} B_{1,\lambda_2\lambda_3} B_{2,\lambda_3\lambda_1} 
 \nonumber \\ 
 & \times
 \int^\beta_0 d\tau_1 \int^\beta_0 d\tau_2~ \Theta(\tau_2-\tau_1) e^{\tau_1 E_{\lambda_3\lambda_2}} e^{\tau_2 E_{\lambda_1\lambda_3}} e^{i\nu_1\tau_1} e^{i\nu_2\tau_2}
 \nonumber \\
 &= \sum'_{\cal P} \sum_{\lambda_1 \lambda_2\lambda_3} P_{\lambda_1}
  A_{\lambda_1\lambda_2} B_{1,\lambda_2\lambda_3} B_{2,\lambda_3\lambda_1} \Bigg \{
   \frac{e^{\beta E_{\lambda_1\lambda_3}} [e^{\beta  E_{\lambda_3\lambda_2} }-1] }{(i\nu_2+E_{\lambda_1\lambda_3})(i\nu_1+E_{\lambda_3\lambda_2})} 
 \nonumber \\ &-
\frac{e^{\beta E_{\lambda_1\lambda_2}}-1}{(i\nu_2+E_{\lambda_1\lambda_3})(i[\nu_1+\nu_2]+E_{\lambda_1\lambda_2})}\Bigg \}~.
\end{align}
Using that $P_\lambda= {e^{-\beta E_\lambda}}/{\cal Z}$,  we can simplify the above relationship as follows  
\begin{align}
 \widetilde \rchi_{A B_1 B_2} (i\nu_1,i\nu_2)  &= -\sum'_{\cal P} \sum_{\lambda_1 \lambda_2\lambda_3} 
 A_{\lambda_1\lambda_2} B_{1,\lambda_2\lambda_3} B_{2,\lambda_3\lambda_1} \Bigg \{
   \frac{ P_{\lambda_3}-P_{\lambda_2} }{(i\nu_2+E_{\lambda_1\lambda_3})(i\nu_1+E_{\lambda_3\lambda_2})} 
   \nonumber \\
   &-
\frac{P_{\lambda_1}-P_{\lambda_2}}{(i\nu_2+E_{\lambda_1\lambda_3})(i[\nu_1+\nu_2]+E_{\lambda_1\lambda_2})} \Bigg \}~.
\end{align}
We can write the fraction in the first line of the previous equation in the following way  
\begin{equation}
\frac{1}{i\nu_1+E_{\lambda_3\lambda_2}} \frac{1}{i\nu_2+E_{\lambda_1\lambda_3}} = \frac{1}{i[\nu_1+\nu_2]+E_{\lambda_1\lambda_2}} \left [\frac{1}{i\nu_1+E_{\lambda_3\lambda_2}} + \frac{1}{i\nu_2+E_{\lambda_1\lambda_3}} \right ]~.
\end{equation}
Using the above identity, we can simplify the expression of $\widetilde \rchi_{A B_1 B_2} (i\nu_1,i\nu_2)$ even more and reach 
\begin{align}
 &\widetilde \rchi_{A B_1 B_2} (i\nu_1,i\nu_2)
= -\sum'_{\cal P} \sum_{\lambda_1 \lambda_2\lambda_3} 
 \frac{A_{\lambda_1\lambda_2} B_{1,\lambda_2\lambda_3} B_{2,\lambda_3\lambda_1}}{i[\nu_1+\nu_2]+E_{\lambda_1\lambda_2}}
 \nonumber \\ &\times
  \left \{
\frac{P_{\lambda_3}-P_{\lambda_2} }{i\nu_1+E_{\lambda_3\lambda_2}} + \frac{P_{\lambda_3}-P_{\lambda_2} }{i\nu_2+E_{\lambda_1\lambda_3}} 
   -
\frac{P_{\lambda_1}-P_{\lambda_2}}{i\nu_2+E_{\lambda_1\lambda_3}}\right \}
\nonumber \\
 &=- \sum'_{\cal P} \sum_{\lambda_1 \lambda_2\lambda_3} 
 \frac{A_{\lambda_1\lambda_2} B_{1,\lambda_2\lambda_3} B_{2,\lambda_3\lambda_1}}{i[\nu_1+\nu_2]+E_{\lambda_1\lambda_2}} \left \{
\frac{P_{\lambda_3}-P_{\lambda_2} }{i\nu_1+E_{\lambda_3\lambda_2}} -\frac{P_{\lambda_1}-P_{\lambda_3}}{i\nu_2+E_{\lambda_1\lambda_3}}
 \right \}~.
\end{align}
After exchanging $\lambda_2$ with $\lambda_3$, we obtain
\begin{equation}
  \widetilde \rchi_{A B_1 B_2} (i\nu_1,i\nu_2) = \sum'_{\cal P} \sum_{\lambda_1 \lambda_2\lambda_3} 
 \frac{A_{\lambda_1\lambda_{3}} B_{1,\lambda_{3}\lambda_{2}} B_{2,\lambda_{2}\lambda_1}}{i[\nu_1+\nu_2]+E_{\lambda_1\lambda_{3}}} \left \{
\frac{P_{\lambda_1}-P_{\lambda_{2}}}{i\nu_2+E_{\lambda_1\lambda_{2}}}
-
\frac{P_{\lambda_{2}}-P_{\lambda_{3}} }{i\nu_1+E_{\lambda_{2}\lambda_{3}}} 
 \right \}~.
\end{equation}
By considering the intrinsic permutation operation, we reach the following compact form for the spectral representation of the second-order response in the Matsubara frequency domain:  
\begin{equation}
   \widetilde \rchi_{A B_1 B_2} (i\nu_1,i\nu_2) = 
   \sum'_{\cal P} \sum_{\lambda_1 \lambda_2\lambda_3}\frac{A_{\lambda_1\lambda_{3}} B_{2,\lambda_{3}\lambda_{2}} B_{1,\lambda_{2}\lambda_1}}{i[\nu_1+\nu_2]+E_{\lambda_1\lambda_{3}}} 
   \left \{
  \frac{P_{\lambda_1\lambda_2}}{i\nu_1+E_{\lambda_1\lambda_{2}}}
-
\frac{P_{\lambda_{2}\lambda_{3}} }{i\nu_2+E_{\lambda_{2}\lambda_{3}}} 
 \right \}~.
\end{equation}
By performing the analytical continuation $i\nu_i \to \omega_i+ i\eta_i$ we obtain the physical (retarded) second-order response function given in 
Eq.~(\ref{eq:chi2_ret}). Once again, we should set $\eta_1=\eta_2$ in order to turn on all the external fields at the same rate in the adiabatic switching-on procedure. 

\subsection{Analytic continuation for higher order nonlinear response functions}

We define the $m$-th order time-ordered response function as follows:
\begin{equation}\label{eq:mth_corr}
\widetilde \rchi_{A B_1 B_2 \dots B_m} (\tau_1,\tau_2,\dots,\tau_m)= \frac{(-1)^m}{m!} \left\langle { \cal T} \hat B_m (-\tau_m) \dots \hat B_2 (-\tau_2) \hat B_1 (-\tau_1) \hat A(0) \right \rangle~,
\end{equation}
where $\tau_i$ denotes an imaginary time. Notice that the ${(-1)^m}/{m!}$ factor is required in order to achieve consistency with the real-time picture, see e.g.~Eq.~(\ref{eq:chi2_ret}). 
The intrinsic permutation symmetry is implicitly taken into account through the presence of the time-ordering operation, ${\cal T}$, and the $1/m!$ pre-factor is there to avoid multiple counting.  A similar relation for the imaginary-time nonlinear correlation function was first reported in Ref.~\cite{Evans_1966}. The response function can be expressed in terms of a ``universal"  kernel $X^{(n)}$  in the following manner:
\begin{align}\label{eq:chi_n}
\widetilde \rchi_{A B_1\dots B_n}(i\nu_1,\dots,i\nu_n) &= \sum'_{\cal P} \sum_{\{\lambda_i\}}   X^{(n)}_{\lambda_1\dots\lambda_{n+1}}(i\nu_1,\dots,i\nu_n) 
\left[A_{\lambda_1\lambda_{n+1}} \Pi^n_{i=1}B_{i,\lambda_{i+1}\lambda_{i}} \right]~.
\end{align}
The universal kernel, which entirely accounts for the frequency dependence of the response, can be evaluated in a recursive manner as described below. 
In the  zero-th order it is given by the statistical occupation factor: 
\begin{align}
X^{(0)}_{\lambda} =  P_{\lambda} = \frac{e^{-\beta E_\lambda}}{\sum_\lambda e^{-\beta E_\lambda}}~.
\end{align}
The first-order case then follows
\begin{align}
X^{(1)}_{\lambda_1\lambda_2}(i\nu_1) =   \frac{X^{(0)}_{\lambda_1} -X^{(0)}_{\lambda_2} }{i\nu_1+E_{\lambda_1\lambda_{2}}}~.
\end{align}
Following the derivation given above regarding the analytical continuation of the second-order response function, we find
\begin{equation}
   X^{(2)}_{\lambda_1\lambda_2\lambda_3}(i\nu_1,i\nu_2) = 
\frac{X^{(1)}_{\lambda_1\lambda_2}(i\nu_1)-X^{(1)}_{\lambda_2\lambda_3}(i\nu_2)}{i[\nu_1+\nu_2]+E_{\lambda_1\lambda_{3}}}~.
\end{equation}
Similarly, for the third order we obtain 
\begin{equation}
   X^{(3)}_{\lambda_1\lambda_2\lambda_3\lambda_4}(i\nu_1,i\nu_2,i\nu_3) = 
\frac{X^{(2)}_{\lambda_1\lambda_2\lambda_3}(i\nu_1,i\nu_2)-X^{(2)}_{\lambda_2\lambda_3\lambda_4}(i\nu_2,i\nu_3)}{i[\nu_1+\nu_2+\nu_3]+E_{\lambda_1\lambda_4}} 
~,
\end{equation}
and thus eventually for the $n$-th order we have
\begin{equation}\label{Recursion}
   X^{(n)}_{\lambda_1\dots\lambda_{n+1}}(i\nu_1,\dots,i\nu_n) = 
\frac{X^{(n-1)}_{\lambda_1\dots\lambda_{n}}(i\nu_1,\dots,i\nu_{n-1})-X^{(n-1)}_{\lambda_2\dots\lambda_{n+1}}(i\nu_2,\dots,i\nu_n)}{i[\nu_1+\dots+\nu_n]+E_{\lambda_1\lambda_{n+1}}} 
~.
\end{equation}
This is the desired recursion relation.
Having the universal $X^{(n)}$ response function, one can calculate the $n$-th order physical response function after properly incorporating the form-factor part shown in the square bracket in Eq.~(\ref{eq:chi_n}) and summing on all degrees of freedoms $\lambda_i$. 
We have already established that, up to second-order, the analytic continuation from time-ordered to causal response is effected by the replacement,  $i\nu_i \to \omega_i+ i\eta_i$.   The recursion relation (\ref{Recursion}) shows that the same procedure will also work for the $n$-th order response if it works for the $n-1$-th response. Thus, we have provided an inductive proof of the analytic continuation procedure {\it at all orders}. Our proof is considerably simpler than the one first reported in Ref.~\cite{Evans_1966}. 

\section{Light-matter interaction: gauge transformation}\label{sec:light_matter}
After the above formal analysis of nonlinear response functions, we now proceed to discuss the issue of gauge invariance in nonlinear response theory, analysing the role of an external electromagnetic field.  
We then generalize the linear Ward identity stemming from gauge invariance and charge conservation 
 to higher-order response functions~\cite{Peskin_and_Schroeder,cheng_sr_2017}.

In the interaction representation, the light-matter interaction can be written in terms of vector, ${\bm {\mathcal A}}({\bm r},t) $, and scalar, $\Phi({\bm r},t)$, potentials in the space-time domain, $({\bm r},t)$:  
\begin{equation}\label{eq:V}
\hat V(t)= - \sum_{\alpha_1} \int \left(\int^1_0 d\lambda~\hat {\cal J}_{\alpha_1}({\bm r},t;
\lambda {\bm  {\mathcal A}})\right) {\cal A}_{\alpha_1}({\bm r},t) d{\bm r} + \int \hat n({\bm r},t) \Phi({\bm r},t) d{\bm r}~,
\end{equation}
where  $\hat {\cal J}_{\alpha_1}({\bm r},t;{\bm  {\mathcal A}})$ is the charge current operator, which, in general, may depend on the electromagnetic vector potential ${\bm  {\mathcal A}}$. 
The integration over the real parameter $\lambda$ in the first term of Eq.~(\ref {eq:V})  guarantees the fundamental relation between the current and the light-matter interaction Hamiltonian:
\begin{equation}
\hat {\cal J}_{\alpha_1}({\bm r},t;{\bm  {\mathcal A}}) =\frac{\delta \hat V(t)}{\delta{\cal A}_{\alpha_1}({\bm r},t)}
\end{equation}
for any value of ${\bm  {\mathcal A}}$.

Note that the time dependence of the operators $\hat V$ and $\hat {\cal J}_\alpha$ originates from two sources: the interaction-picture of the time evolution, see Eq.~(\ref{eq:interaction_picture}), and the explicit time dependence of the external potentials. The time dependence of the charge-density operator $\hat n$ stems from the interaction-picture of the time evolution. The charge-current operator in the interaction picture is given by
\begin{align}
& \hat {\cal J}_{\alpha_1} ({\bm r},t;{\bm  {\mathcal A}})  = \hat j_{\alpha_1}({\bm r},t) +\sum_{\alpha_2} \int d{\bm r'} \hat \kappa_{\alpha_1\alpha_2}({\bm r},{\bm r}',t) {\cal A}_\beta({\bm r}',t)
\nonumber\\
&+\sum_{\alpha_2\alpha_3} \int d{\bm r}' \int d{\bm r}'' \hat \xi_{\alpha_1\alpha_2\alpha_3}({\bm r},{\bm r}',{\bm r}'',t) {\cal A}_{\alpha_2}({\bm r}',t) {\cal A}_{\alpha_3}({\bm r}'',t)
\nonumber\\
&+\sum_{\alpha_2\alpha_3\alpha_4} \int d{\bm r}' \int d{\bm r}'' \int d{\bm r}''' \hat \zeta_{\alpha_1\alpha_2\alpha_3\alpha_4}({\bm r},{\bm r}',{\bm r}'',t) {\cal A}_{\alpha_2}({\bm r}',t) {\cal A}_{\alpha_3}({\bm r}'',t) {\cal A}_{\alpha_4}({\bm r}''',t)
\nonumber \\
&+\dots~,
\end{align}
where $\hat {\cal H}_{\rm tot} = \hat {\cal H}+ \hat V$.
Here, the one-photon current component in the interaction picture is given by 
\begin{align}
\hat j_{\alpha_1} ({\bm r},t) &= \hat {\cal J}_{\alpha_1} ({\bm r},t;{\bm {\mathcal A}})|_{{\bm {\mathcal A}}  \to 0}= 
-
e^{i t \hat{\cal H}}
\frac{\delta \hat {\cal H}_{\rm tot}}{\delta {\cal A}_\alpha({\bm r},t)}  \Big |_{{\bm {\mathcal A}}  \to 0}
e^{-i t \hat{\cal H}}
~.
\end{align}
For the two-photon current coupling we have
\begin{align}
\hat \kappa_{\alpha_1\alpha_2} ({\bm r},{\bm r}',t) &=
\frac{\delta \hat {\cal J}_{\alpha_1} ({\bm r},t;{\bm {\mathcal A}})}{\delta {\cal A}_{\alpha_2}({\bm r}',t)}  \Big |_{{\bm {\mathcal A}}  \to 0}= -
e^{i t \hat{\cal H}}
\frac{\delta^2 \hat {\cal H}_{\rm tot}}{\delta {\cal A}_{\alpha_1}({\bm r},t) \delta {\cal A}_{\alpha_2}({\bm r}',t)}  \Big |_{{\bm {\mathcal A}}  \to 0}
e^{-i t \hat{\cal H}}
~.
\end{align}
Similarly, three- and four-photon current couplings read as following:
\begin{align}
\hat \xi_{\alpha_1\alpha_2\alpha_3}({\bm r},{\bm r}',{\bm r}'',t)  &
=
\frac{1}{2!}\frac{\delta^2 \hat {\cal J}_{\alpha_1} ({\bm r},t;{\bm {\mathcal A}})}{\delta {\cal A}_{\alpha_2}({\bm r}',t) \delta {\cal A}_{\alpha_3}({\bm r}'',t)}  \Big |_{{\bm {\mathcal A}}  \to 0}
\nonumber\\& = 
-
e^{i t \hat{\cal H}}
\frac{1}{2!}\frac{\delta^3 \hat {\cal H}_{\rm tot}}{\delta {\cal A}_{\alpha_1}({\bm r},t) \delta {\cal A}_{\alpha_2} ({\bm r}',t) \delta {\cal A}_{\alpha_3} ({\bm r}'',t)}  \Big |_{{\bm {\mathcal A}}  \to 0}
e^{-i t \hat{\cal H}}
 \end{align}
and
\begin{align}
&\hat \zeta_{\alpha_1\alpha_2\alpha_3\alpha_4}({\bm r},{\bm r}',{\bm r}'',{\bm r}''',t)  
=
\frac{1}{3!}\frac{\delta^3 \hat {\cal J}_{\alpha_1}({\bm r},t;{\bm {\mathcal A}})}{\delta {\cal A}_{\alpha_2}({\bm r}',t) \delta {\cal A}_{\alpha_3} ({\bm r}'',t) \delta {\cal A}_{\alpha_4} ({\bm r}''',t)}  \Big |_{{\bm {\mathcal A}}  \to 0}
\nonumber\\& = -
e^{i t \hat{\cal H}}
\frac{1}{3!}\frac{\delta^4 \hat {\cal H}_{\rm tot}}{\delta {\cal A}_{\alpha_1}({\bm r},t) \delta {\cal A}_{\alpha_2} ({\bm r}',t) \delta {\cal A}_{\alpha_3} ({\bm r}'',t)  \delta {\cal A}_{\alpha_4} ({\bm r}''',t)}  \Big |_{{\bm {\mathcal A}}  \to 0}
e^{-i t \hat{\cal H}}
~.
 \end{align}
It is common to dub $\hat j_{\alpha_1}$ ``{\it paramagnetic}" current operator  and $\hat \kappa_{\alpha_1\alpha_2} $,  $\hat \xi_{\alpha_1\alpha_2\alpha_3}$,  and $\hat \zeta_{\alpha_1\alpha_2\alpha_3\alpha_4}$ ``{\it multi-photon}" current operators (which can have either a {\it diamagnetic} or a {\it paramagnetic} contribution to total current depending on the ground-state phase and the order of perturbation; jargon stemming from the theory of superconductivity~\cite{Schrieffer}). 
Note, finally, that the time dependence of $\hat j_{\alpha_1}({\bm r},t)$ (and all multi-photon current operators) stems from the interaction-picture of the time evolution, see Eq.~(\ref{eq:interaction_picture}).

Formally, we have the following perturbative series for the macroscopic charge density and current
\begin{align}
&N({\bm r},t) = \left  \langle \hat n({\bm r},t) \right \rangle + N^{(1)}({\bm r}, t) + N^{(2)}({\bm r}, t) +N^{(3)}({\bm r}, t) +\dots ~, \label{eq:N_series}
\\
&J_\ell({\bm r},t) = \left  \langle \hat j_\ell({\bm r},t) \right \rangle + J^{(1)}_\ell({\bm r}, t) + J^{(2)}_\ell({\bm r}, t) +J^{(3)}_\ell({\bm r}, t) +\dots~. \label{eq:J_series}
\end{align}
We now apply a gauge transformation to obtain gauge-invariant relations for the first- and second-order density and current. 
The gauge transformation is defined by  the following map 
\begin{align}
 {\bm {\mathcal A}}'({\bm r},t) = {\bm {\mathcal A}}({\bm r},t) + {\bm \nabla} \Lambda ({\bm r},t) ~~,
~~
 \Phi'({\bm r},t) = \Phi({\bm r},t) -\frac{\partial \Lambda({\bm r},t)}{\partial t}~,
\end{align}
where $ \Lambda ({\bm r},t) $ is an arbitrary smooth function of ${\bm r}$ and $t$.
Similarly, in Fourier space: 
\begin{align}\label{eq:gauge_transformation}
{\cal A}'_{\alpha}({\bm q},\omega) = {\cal A}_{\alpha}({\bm q},\omega) +  i q_\alpha  \Lambda ({\bm q},\omega)~~,
~~
\Phi'({\bm q},\omega) = \Phi({\bm q},\omega) + i \omega \Lambda ({\bm q},\omega)~.
\end{align}
The gauge transformation preserves the electric and  magnetic fields, 
\begin{equation}\label{EMFields}
{\bm E}({\bm r},t)=-{\bm \nabla } \Phi({\bm r},t) - \frac{\partial {\bm  {\mathcal A}}({\bm r},t)}{\partial t}~~,~~
{\bm B}({\bm r},t)= {\bm \nabla }  \times {\bm  {\mathcal A}}({\bm r},t)~,
\end{equation}
where we have set $c=1$ for the speed of light in vacuum. In Fourier space, the  field components read
\begin{equation}\label{eq:E_q_omega}
E_\alpha({\bm q},\omega) = -i  q_\alpha \Phi({\bm q},\omega) +i\omega  {\cal A}_\alpha({\bm q},\omega)~~,~~
B_\alpha({\bm q},\omega)= i \sum_{\beta\gamma}\epsilon_{\alpha\beta\gamma} q_\beta    {\mathcal A}_\gamma({\bm q},\omega)~,
\end{equation}
where $\epsilon_{\alpha\beta\gamma} $ is the Levi-Civita symbol.  From now on, we drop the imaginary part of $\omega$ in the notation but keep in mind that $\omega \equiv \omega+i0^+$. 

According to gauge invariance, the difference between physical observables calculated in two arbitrary gauges must be zero. This implies: $\delta {N}^{(n)} = {N}^{' (n)} - {N}^{(n)} = 0$ and $\delta {J}^{(n)}_\ell = {J}^{' (n)}_\ell - {J}^{(n)}_\ell = 0$~. 

\section{Linear response theory: gauge invariance }\label{sec:linear}
By truncating the perturbative series up to linear order in the external field (see~\ref{app:rt_o1_and_o2}), we reach the following formal relations for the linear charge density and current:
\begin{align} \label{eq:n1_j1}
N^{(1)}({\bm q}, \omega) &=  \sum_{{\bm q}'}
\left \{ 
\rchi_{nn}({\bm q},{\bm q}',\omega) \Phi({\bm q}', \omega)
- \sum_{\alpha} \rchi_{n j_\alpha}({\bm q},{\bm q}',\omega) {\cal A}_\alpha({\bm q}', \omega) 
\right \}, 
\\
  J^{(1)}_\ell ({\bm q},\omega) &= \sum_{{\bm q}'}
 \Bigg \{ 
\sum_{\alpha} 
 \Big [ 
\left \langle \hat \kappa_{\ell\alpha}({\bm q}, {\bm q}') \right \rangle   
- \rchi_{ j_{\ell} j_{\alpha} } ({\bm q},{\bm q}',\omega) 
\Big ]
{\cal A}_{\alpha}({\bm q}',\omega)
\nonumber\\& +\rchi_{j_{\ell} n } ({\bm q},{\bm q}',\omega) \Phi({\bm q}',\omega) 
\Bigg \}. 
\end{align}
Note that our system is not  assumed to be translationally invariant and consequently ${\bm q}'$ and ${\bm q}$ are two independent variables. 

According to gauge invariance and using Eq.~(\ref{eq:gauge_transformation}), the difference between the physical charge densities and currents calculated in two different gauges must be identically zero:
\begin{align} 
\delta N^{(1)}({\bm q}, \omega) &=
i \sum_{{\bm q}'}\left \{  
\omega \rchi_{nn}({\bm q},{\bm q}',\omega) - \sum_{\alpha}\rchi_{n j_\alpha}({\bm q},{\bm q}',\omega) q'_\alpha  \right \} \Lambda({\bm q}', \omega)=0,
\nonumber \\
\hspace{-1cm}
\delta J^{(1)}_\ell({\bm q},\omega) &=
i\sum_{{\bm q}'} \Bigg \{ \sum_{\alpha}  
\left [ \left \langle \hat \kappa_{\ell\alpha}({\bm q}, {\bm q}') \right \rangle - \rchi_{ j_{\ell} j_{\alpha} } ({\bm q},{\bm q}',\omega) 
\right ] q'_{\alpha} 
+ \omega  \rchi_{j_{\ell} n } ({\bm q},{\bm q}',\omega) \Bigg \}
\nonumber\\& \times \Lambda({\bm q}',\omega)  = 0.
\end{align}
In order to fulfill these constraints, the following gauge-invariance identities must be satisfied  
\begin{align}\label{eq:gauge_invariance_identity_n1_j1}
\rchi_{nn}({\bm q},{\bm q}',\omega) &= \sum_{\alpha} \frac{q'_\alpha }{ \omega }\rchi_{n j_\alpha}({\bm q},{\bm q}',\omega)~,
\nonumber \\
\rchi_{j_{\ell} n } ({\bm q},{\bm q}',\omega) &=  \sum_{\alpha} \frac{q'_{\alpha}}{\omega} \left [ 
 \rchi_{ j_{\ell} j_{\alpha} } ({\bm q},{\bm q}',\omega) - \left \langle \hat \kappa_{\ell\alpha}({\bm q}, {\bm q}') \right \rangle \right ]~.
\end{align}
After implementing the above identities in Eq.~(\ref{eq:n1_j1}) and using Eq.~(\ref{eq:E_q_omega}), we can write down the following gauge-invariant form of the linear density and current:
\begin{align}
N^{(1)}({\bm q}, \omega) &=  - \sum_{{\bm q}'}  \sum_{\alpha}\rchi_{n j_\alpha}({\bm q},{\bm q}',\omega) \frac{E_\alpha({\bm q}', \omega)}{i\omega}~, 
\\
J^{(1)}_\ell({\bm q},\omega) &= - \sum_{{\bm q}'} \sum_{\alpha}  
\left [ 
 \rchi_{ j_{\ell} j_{\alpha} } ({\bm q},{\bm q}',\omega)   
 -
 \left \langle \hat \kappa_{\ell\alpha}({\bm q}, {\bm q}') \right \rangle \right ]
 \frac{E_{\alpha}({\bm q}',\omega)}{i\omega}~. \label{eq:gauge_ invariant_j1}
\end{align}
The surprising absence of the magnetic field in these equations is explained by noting that the finite frequency components of the magnetic field are connected to the electric field by the two ``internal"  Maxwell equations, e.g. Faraday's law ${\bm q}\times {\bm E}({\bm q},\omega) = \omega {\bm B}({\bm q},\omega)$ and ${\bm q}\cdot{\bm B}({\bm q},\omega)=0$, which follow from Eq.~(\ref{eq:E_q_omega}). If we expand the conductivity up to linear order in $\bm q$ we can generate terms proportional to the magnetic field.

Because of the above relation for the first-order current, we have the following expression for the linear conductivity:
\begin{equation}
\sigma^{(1)}_{\ell\alpha} ({\bm q},{\bm q}',\omega) = - \frac{\rchi_{ j_{\ell} j_{\alpha} } ({\bm q},{\bm q}',\omega) - \left \langle \hat \kappa_{\ell\alpha}({\bm q}, {\bm q}') \right \rangle }{i\omega}~.
\end{equation}
Utilizing the diamagnetic sum rule for a longitudinal external field $ {\bm E}({\bm q}',\omega) || {\bm q'} $, we have (see Ref.~\cite{Giuliani_and_Vignale} and also \ref{app:linear_diamagnetic})
\begin{equation}\label{eq:diamagnetic}
\text{For longitudinal $\hat j_\alpha$:~~~} \left \langle \hat \kappa_{\ell\alpha}({\bm q}, {\bm q}') \right \rangle =   \lim_{\omega\to0}  \rchi_{ j_{\ell} j_{\alpha} } ({\bm q},{\bm q}',\omega)~.
\end{equation}
For a transverse field ${\bm E}({\bm q}',\omega) \perp {\bm q'} $ the total static current response is finite and therefore the diamagnetic sum rule is valid only at  ${\bm q}'=0$:
\begin{equation}\label{eq:diamagnetic}
\text{For transverse $\hat j_\alpha$:~~~}  \left \langle \hat \kappa_{\ell\alpha}({\bm q}, {\bm q}'=0) \right \rangle =  \lim_{\bm q'\to0} \lim_{\omega\to0}  \rchi_{ j_{\ell} j_{\alpha} } ({\bm q},{\bm q}',\omega)~.
\end{equation}
The diamagnetic sum rule identity is explicitly proven in~\ref{app:linear_diamagnetic} by using the well-known Kubo identity~\cite{Louie_and_Cohen,Clark_and_Derrick} and the continuity relation. This cancellation clarifies the importance of the diamagnetic component of the current operator, i.e. $\sum_\alpha \hat \kappa_{\ell\alpha} {\cal A}_\alpha$, in the linear conductivity. Finally, we recall that in the homogeneous (translational invariant) electron liquid we have $\bm q=\bm q'$. 
In the clean system, the Drude divergence of the dc limit, $\omega\to0$, is achieved in the reverse order of limits that is first setting $\bm q\to 0$ and then $\omega\to0$. In this case, the paramagnetic response function vanishes, $\lim_{\omega\to 0} \lim_{\bm q \to 0} \rchi_{j_\ell j_\alpha} \to 0$, in a perfectly homogeneous electron liquid owing to momentum conservation~\cite{Giuliani_and_Vignale}. As a consequence, the diamagnetic contribution leads to the Drude divergence at zero frequency.

The continuity constitution laws for the two- and three-photon diamagnetic current couplings read (see \ref{app:diamagnetic_continuity}):
\begin{align}
[\hat j_{\beta}({\bm r'}),\hat n({\bm r})] = \sum_\alpha (-i\partial_{r_\alpha}) \hat \kappa_{\alpha\beta}({\bm r},{\bm r}')~,
\end{align}
and
\begin{align}
\frac{1}{2}[[ \hat j_{\gamma}({\bm r}''),\hat n({\bm r}')],\hat n({\bm r})] = \sum_{\alpha\beta} (-i\partial_{r'_\beta})(-i\partial_{r_\alpha}) \hat \xi_{\alpha\beta\gamma}({\bm r},{\bm r}',{\bm r}'')~.
\end{align}
 A similar strategy can be employed to resolve a connection between the equal-time nested commutation of the  paramagnetic current and the density operator with higher-order $n$-photon diamagnetic current operators. Let us focus on the two-photon diamagnetic current coupling, which, in Fourier space, obeys
\begin{align}
 [\hat j_{\beta}(-\bm q') ,\hat n({\bm q})] = \sum_\alpha q_\alpha \hat \kappa_{\alpha\beta}({\bm q},{\bm q}')~.
\end{align}
It is straightforward to evaluate $[\hat j_{\beta}(-\bm q') ,\hat n({\bm q})]$ and obtain (see \ref{app:diamagnetic_continuity})
 \begin{align}
 \sum_\alpha q_\alpha \hat \kappa_{\alpha\beta}({\bm q},{\bm q}') = e^2 v_{\beta}(-{\bm q}) \hat n(\bm q-\bm q')~.
 \end{align} 
Notice that $e>0$ is the fundamental electric charge and $v_\beta({\bm q}) = \partial_{q_\beta} {\cal H}({\bm q})$ is the velocity operator.

\section{Second-order response theory: gauge invariance}\label{sec:second}
In order to obtain the second-order density and current, we need to keep all terms up to quadratic order in the external field. The details of this calculation are available in~\ref{app:rt_o1_and_o2} in which we start from the position space and real time representation. 
Carrying out lengthy but straightforward calculations (see~\ref{app:rt_o1_and_o2}), 
we reach the following relation for the second-order density in the Fourier representation:   
\begin{align} \label{eq:N2}
 N^{(2)}({\bm q}, \omega_\Sigma) &=  \sum_{{\bm q}_1 {\bm q}_2} 
\Big \{  
\sum_{\alpha_1\alpha_2 } \Pi_{n;\alpha_1\alpha_2 } ({\bm Q}_{12},\Omega_{12}) {\cal A}_{\alpha_1}({\bm q}_1, \omega_1) {\cal A}_{\alpha_2} ({\bm q}_2, \omega_2)
\nonumber\\
&-\sum_{\alpha_2 } \Pi_{n; n \alpha_2}({\bm Q}_{12},\Omega_{12}) \Phi({\bm q}_2, \omega_1) {\cal A}_{\alpha_2} ({\bm q}_2, \omega_2)
\nonumber \\
&+   \Pi_{n;nn}  ({\bm Q}_{12},\Omega_{12})\Phi({\bm q}_1, \omega_1)  \Phi({\bm q}_2, \omega_2)
 \Big \}~,
\end{align}
where we have introduced the shortand $({\bm q},{\bm q}_1,{\bm q}_2,\omega_1,\omega_2)=({\bm Q}_{12},\Omega_{12})$.
Notice that $ \Pi_{n; n n } ({\bm Q}_{12},\Omega_{12}) = \sum^\prime_{\cal P} \rchi_{n n n} ({\bm Q}_{12},\Omega_{12})$, where the $ \Pi_{n; \alpha_1\alpha_2}$ and $ \Pi_{n; n\alpha_2 }$ response functions are defined by
\begin{align}
 & \Pi_{n; \alpha_1\alpha_2} ({\bm Q}_{12},\Omega_{12}) = 
 \sum^\prime_{\cal P} \left \{
 \rchi_{n j_{\alpha_1} j_{\alpha_2}} ({\bm Q}_{12},\Omega_{12}) 
 - 
 \rchi_{n \kappa_{\alpha_1\alpha_2}} ({\bm Q}_{12},\Omega_{12})\right \}~, 
\nonumber\\ & 
\Pi_{n; n\alpha_2 } ({\bm Q}_{12},\Omega_{12}) =
  \rchi_{n n j_{\alpha_2}} ({\bm Q}_{12},\Omega_{12})
 +\rchi_{n j_{\alpha_2} n} ({\bm Q}_{21},\Omega_{21})~.  
\end{align}
Note that the second-order correlation function $\rchi_{ABC}$ is defined generically in Eq.~(\ref{eq:chi2}) and (\ref{eq:chi2_FQ}); see also Eq.~(\ref{eq:B8}) and Eq.~(\ref{eq:B9}).
Here, $\sum'_{\cal P} $ stands for the intrinsic permutation symmetry of each correlation function with respect to the dummy variables~\cite{Butcher_and_Cotter}. Notice that in our notation $`` \Pi_{A;B_1B_2\dots} F_1 F_2 \dots"$   
denotes the macroscopic value of the $``\hat A"$ operator in response to the $``F_i"$ external fields, which are coupled to the $``\hat B_i"$ operators.

Carrying out the gauge transformation (\ref{eq:gauge_transformation}) and following similar steps to those described in the linear-response case, we reach the following relation for the difference of the second-order densities between the two gauges: 
\begin{align}\label{eq:deltaN2=0}
\delta N^{(2)}({\bm q},\omega_\Sigma)  &=  \sum_{{\bm q}_1}   
\Bigg \{
i \left[\sum_{{\bm q}_2}K_1 ({\bm Q}_{12},\Omega_{12}; { \bm {\mathcal A}},\Phi) \right]\Lambda ({\bm q}_1, \omega_1) 
\nonumber\\& -
 \sum_{{\bm q}_1,{\bm q}_2}   K_2 ({\bm Q}_{12},\Omega_{12}) \Lambda({\bm q}_1,\omega_1) \Lambda({\bm q}_2,\omega_2) 
 \Bigg \}~.
\end{align}
Here, 
\begin{align}
K_2({\bm Q}_{12},\Omega_{12}) &=
\sum_{\alpha_1\alpha_2 } \Pi_{n;\alpha_1\alpha_2 } ({\bm Q}_{12},\Omega_{12}) 
q_{1,\alpha_1} q_{2,\alpha_2}  
- \sum_{\alpha_2 } \Pi_{n; n \alpha_2}({\bm Q}_{12},\Omega_{12})
 \omega_1 q_{2,\alpha_2}  
 \nonumber\\
&+
\Pi_{n;nn}  ({\bm Q}_{12},\Omega_{12})
\omega_1 \omega_2
 \end{align}
 and 
 \begin{align}\label{eq:K1}
K_1 ({\bm Q}_{12},\Omega_{12}; {\bm{\mathcal A}},\Phi)& =
\sum_{\alpha_1\alpha_2 } 
\Big [
\Pi_{n;\alpha_1\alpha_2 } ({\bm Q}_{12},\Omega_{12}) +\Pi_{n;\alpha_2 \alpha_1 } ({\bm Q}_{21},\Omega_{21}) 
\Big ]
\nonumber\\& \times q_{1,\alpha_1}  {\cal A}_{\alpha_2} ({\bm q}_2, \omega_2)  
\nonumber\\ 
&-
\sum_{\alpha_2 } 
\Big [ 
\Pi_{n; n \alpha_2}({\bm Q}_{12},\Omega_{12})  \omega_1  {\cal A}_{\alpha_2} ({\bm q}_2, \omega_2)
 \nonumber \\ &
 + 
 \Pi_{n; n \alpha_2}({\bm Q}_{21},\Omega_{21}) \Phi({\bm q}_2, \omega_2) q_{1,\alpha_2}
 \Big ]
\nonumber\\
&+
\Big [ 
\Pi_{n;nn}  ({\bm Q}_{12},\Omega_{12}) 
+
\Pi_{n;nn}  ({\bm Q}_{21},\Omega_{21})  
\Big ] 
\omega_1  \Phi({\bm q}_2, \omega_2)~.
 \end{align}
Notice that in the argument of $K_1$ we use ${\bm{\mathcal  A}}$ and $\Phi$ to point out the functional dependence of $K_1$ on the external fields. 
For gauge invariance to hold at second-order perturbation theory, the quantity $\delta N^{(2)}({\bm q},\omega_\Sigma)$ must vanish identically.
Since $\delta N^{(2)}({\bm q},\omega_\Sigma)=0$ must be true for any $\Lambda$, we need to have  $\sum_{\bm q_2}K_1=0$ and $K_2=0$ separately.
The condition $K_2=0$ implies 
\begin{align}\label{eq:quad_n_Lambda}
& \sum_{\alpha_1\alpha_2} \Pi_{n;\alpha_1\alpha_2} ({\bm Q}_{12},\Omega_{12}) 
q_{1,\alpha} q_{2,\alpha_2}  
-\sum_{\alpha_2} \Pi_{n; n \alpha_2}({\bm Q}_{12},\Omega_{12})
 \omega_1 q_{2,\alpha_2}  
\nonumber\\& +\Pi_{n;nn}  ({\bm Q}_{12},\Omega_{12})
\omega_1 \omega_2 =0~.
\end{align}

Replacing Eq.~(\ref{eq:quad_n_Lambda}) in Eq.~(\ref{eq:N2}) we obtain a new relation for the second-order density:
\begin{align}\label{eq:N22}
N^{(2)}({\bm q}, \omega_\Sigma) &=   \sum_{{\bm q}' {\bm q}''} 
\sum_{\alpha_1\alpha_2 } \Pi_{n;\alpha_1\alpha_2 } ({\bm Q}_{12},\Omega_{12}) \frac{ E_{\alpha_1}({\bm q}_1, \omega_1) } {i\omega_1} \frac{ E_{\alpha_2} ({\bm q}_2, \omega_2)}{i\omega_2}
\nonumber\\
&+   \sum_{{\bm q}' {\bm q}''} 
\Big \{
 \sum_{\alpha_2 } U_{\alpha_2}({\bm Q}_{12},\Omega_{12})  E_{\alpha_2} ({\bm q}_2, \omega_2)
\Big \}
 \frac{ \Phi({\bm q}_1, \omega_1)}{i\omega_2 \omega_1}~,
\end{align}
where
$U_{\alpha_2}({\bm Q}_{12},\Omega_{12})  = 2 \sum_{\alpha_1 }  \Pi_{n;\alpha_1\alpha_2 } ({\bm Q}_{12},\Omega_{12})  q_{1,\alpha_1}   
-  \Pi_{n; n \alpha_2}({\bm Q}_{12},\Omega_{12})  \omega_1$.
From Eq.~(\ref{eq:N22}), it is evident that there is still  a dependence on the scalar potential that makes the new representation of $N^{(2)}$ gauge dependent.
This is because we  should also satisfy $ \sum_{\bm q_2} K_1=0$ to reach a fully gauge invariant expression for $N^{(2)}$. However, instead of using $ \sum_{\bm q_2} K_1=0$, we perform another gauge transformation on Eq.~(\ref{eq:N22}):
\begin{align} \label{eq:previous_second_way}
\delta N^{(2)}({\bm q}, \omega_\Sigma)=
-  \sum_{{\bm q}_1 {\bm q}_2} 
\Big \{ 
 \sum_{\alpha_2 }  U_{\alpha_2}({\bm Q}_{12},\Omega_{12})  E_{\alpha_2} ({\bm q}_2, \omega_2)
\Big \} 
 \frac{\Lambda ({\bm q}_1, \omega_1)}{\omega_2} =0~.
\end{align}
From Eq.~(\ref{eq:previous_second_way}) we conclude that
\begin{equation}\label{eq:UE=0}
\sum_{\bm q_2} \sum_{\alpha_2 } U_{\alpha_2}({\bm Q}_{12},\Omega_{12})  E_{\alpha_2} ({\bm q}_2, \omega_2) =0~.
\end{equation}
By replacing Eq.~(\ref{eq:UE=0}) in Eq.~(\ref{eq:N22}), we finally obtain a gauge-invariant expression for the second-order density: 
\begin{equation} \label{eq:gauge_invariance_N2}
N^{(2)}({\bm q}, \omega_\Sigma) =    \sum_{{\bm q}_1 {\bm q}_2} 
\sum_{\alpha_1\alpha_2 } \Pi_{n;\alpha_1\alpha_2 } ({\bm Q}_{12},\Omega_{12}) \frac{ E_{\alpha_1}({\bm q}_1, \omega_1) } {i\omega_1} \frac{ E_{\alpha_2} ({\bm q}_2, \omega_2)}{i\omega_2}~.
\end{equation}
Note that, since Eq.~(\ref{eq:UE=0}) is valid for an arbitrary electric field, it must be 
\begin{equation}\label{eq:U=0}
 U_{\alpha_2}({\bm Q}_{12},\Omega_{12}) =0~.
\end{equation}
By plugging Eq.~(\ref{eq:U=0}) in Eq.~(\ref{eq:quad_n_Lambda}), we find
\begin{equation}\label{eq:gauge_invarince_identity_n2}
 \Pi_{n; n n} ({\bm Q}_{12},\Omega_{12})  
= \sum_{\alpha_1\alpha_2}\Pi_{n; \alpha_1 \alpha_2 } ({\bm Q}_{12},\Omega_{12})  \frac{ q_{1,\alpha_1} q_{2,\alpha_2}}{\omega_1\omega_2 }~.
\end{equation}
There is a simpler approach to prove the above identity. Using the scalar potential gauge, we can write the electric field as $E_\alpha ({\bm q},\omega) = -iq_\alpha \Phi({\bm q},\omega)$. The same electric field is obtained by using a longitudinal vector potential ${\cal A}_\alpha({\bm q},\omega)  = -(q_\alpha/\omega) \Phi({\bm q},\omega)$. The second-order density fluctuation calculated in the two gauges must be equal owing to gauge invariance. Therefore, we obtain
\begin{align}\label{eq:N2_simpel_way}
N^{(2)}({\bf q},\omega_\Sigma) &=\Pi_{nnn} \Phi({\bf q}_1,\omega_1) \Phi({\bf q}_2,\omega_2)= \sum_{\alpha_1\alpha_2} \Pi_{n \alpha_1\alpha_2} {\cal A}_{\alpha_1}({\bf q}_1,\omega_1) {\cal A}_{\alpha_2}({\bf q}_2,\omega_2)
\nonumber\\ 
&=\sum_{\alpha_1\alpha_2} \Pi_{n \alpha_1\alpha_2} (-q_{1,\alpha_1}\Phi({\bf q}_1,\omega_1)/\omega_1) (-q_{2,\alpha_2}\Phi({\bm q}_2,\omega_2)/\omega_2)~.
\end{align}
Accordingly, we find the same identity given in Eq.~(\ref{eq:gauge_invarince_identity_n2}) after cancelling the scalar potential from the above relation.

The above relation is an identity which holds true to ensure the gauge invariance of the second-order density $N^{(2)}({\bm q}, \omega_\Sigma)$. By using 
Eqs.~(\ref{eq:quad_n_Lambda}) and~(\ref{eq:U=0}) and considering the intrinsic permutation symmetry of $\Pi_{n;\alpha_1\alpha_2}$, one can prove that $ \sum_{\bm q_2} K_1=0$ is also fulfilled. This proof is reported in~\ref{app:K1L1}.
\par
We now proceed to obtain a gauge-invariance relation for the second-order current.  The second-order current in the frequency and wave-vector domain can be written as follows (see~\ref{app:rt_o1_and_o2})
\begin{align}\label{eq:j2}
\hspace{-1cm} 
J^{(2)}_\ell({\bm q}, \omega_\Sigma) &=
 \sum_{{\bm q}_1 {\bm q}_2} 
\Big \{  
\sum_{\alpha_1\alpha_2} 
\Pi_{\ell; \alpha_1\alpha_2} ({\bm Q}_{12},\Omega_{12})  
 {\cal A}_{\alpha_1}({\bm q}_1,\omega_1) {\cal A}_{\alpha_2}({\bm q}_2,\omega_2)
\nonumber\\
&-\sum_{\alpha_2} 
\Pi_{\ell; n \alpha_2} ({\bm Q}_{12},\Omega_{12}) 
\Phi({\bm q}_1,\omega_1) {\cal A}_{\alpha_2}({\bm q}_2,\omega_2)
\nonumber \\
&+ \Pi_{\ell; n n } ({\bm Q}_{12},\Omega_{12})  
\Phi({\bm q}_1,\omega_1) \Phi({\bm q}_2,\omega_2)
\Big \}~,
\end{align}
where $\Pi_{\ell; n n} ({\bm Q}_{12},\Omega_{12})  =  \sum'_{\cal P} \rchi_{j_\alpha n n} ({\bm Q}_{12},\Omega_{12})$ and 
\begin{align}
\Pi_{\ell; \alpha_1 \alpha_2} ({\bm Q}_{12},\Omega_{12})  &=
\sum'_{\cal P} 
\Big [\rchi_{j_\ell j_{\alpha_1} j_{\alpha_2}} ({\bm Q}_{12},\Omega_{12})   
- \rchi_{ j_\ell \kappa_{\alpha_1\alpha_2} } ({\bm q},{\bm q}_1,{\bm q}_2,\omega_1+\omega_2) 
\nonumber  \\ &- 
 \rchi_{\kappa_{\ell \alpha_2}  j_{\alpha_1} } ({\bm q},{\bm q}_1,{\bm q}_2,\omega_1) 
+ 
 \left \langle\hat \xi_{\ell\alpha_1\alpha_2}({\bm q}, {\bm q}_1,{\bm q}_2) \right \rangle  
\Big ]~,
\\
\Pi_{\ell; n \alpha_2} ({\bm Q}_{12},\Omega_{12}) & = 
 \rchi_{j_\ell n j_{\alpha_2}} ({\bm Q}_{12},\Omega_{12})
 + \rchi_{j_\ell j_{\alpha_2} n} ({\bm Q}_{21},\Omega_{21}) 
\nonumber \\&
 - \sum'_{\cal P} \rchi_{\kappa_{\ell\alpha_2} n} ({\bm q},{\bm q}_2,{\bm q}_1,\omega_1)~.
\end{align}
For the following two correlation functions, one needs to be careful in the symmetrization process, which should be carried out as follows
\begin{align}
\sum'_{\cal P}  \rchi_{\kappa_{\ell\alpha_2}  j_{\alpha_1} } ({\bm q},{\bm q}_1,{\bm q}_2,\omega_1) & = \frac{1}{2} 
\left [
 \rchi_{\kappa_{\ell\alpha_2}  j_{\alpha_1} } ({\bm q},{\bm q}_1,{\bm q}_2,\omega_1) 
 + \rchi_{\kappa_{\ell\alpha_1}  j_{\alpha_2} } ({\bm q},{\bm q}_2,{\bm q}_1,\omega_2) 
\right ]~,
\\
\sum'_{\cal P} \rchi_{\kappa_{\ell\alpha_2} n} ({\bm q},{\bm q}_2,{\bm q}_1,\omega_1) &= \frac{1}{2} 
\left [ \rchi_{\kappa_{\ell \alpha_2} n} ({\bm q},{\bm q}_2,{\bm q}_1,\omega_1)
+ \rchi_{\kappa_{\ell \alpha_2} n} ({\bm q},{\bm q}_1,{\bm q}_2,\omega_2) \right]~.
\end{align}

After performing the gauge transformation, we reach the following relation for the second-order gauge-induced current change, $\delta J^{(2)}_\ell ({\bm q}, \omega_\Sigma)$: 
\begin{align}\label{eq:dj2}
\delta J^{(2)}_\ell ({\bm q}, \omega_\Sigma) &=
  \sum_{{\bm q}_1 }  
i \left[\sum_{{\bm q}_2 }   L_{1} ({\bm Q}_{12},\Omega_{12}; { \bm {\mathcal A}},\Phi) \right]\Lambda ({\bm q}_1,\omega_1)
\nonumber \\
&-
  \sum_{{\bm q}_1 {\bm q}_2 }   L_{2} ({\bm Q}_{12},\Omega_{12}) \Lambda ({\bm q}_1,\omega_1)  \Lambda({\bm q}_2,\omega_2) 
= 0~,
\end{align}
where 
\begin{align}
 L_{2} ({\bm Q}_{12},\Omega_{12}) &= \sum_{\alpha_1\alpha_2}\Pi_{\ell; \alpha_1 \alpha_2} ({\bm Q}_{12},\Omega_{12})  
 q_{1,\alpha_1} q_{2,\alpha_2}\
- \sum_{\alpha_2} \Pi_{\ell; n \alpha_2} ({\bm Q}_{12},\Omega_{12}) \omega_1   q_{2,\alpha_2}
\nonumber \\
&+   \Pi_{\ell; n n} ({\bm Q}_{12},\Omega_{12})  \omega_1\omega_2 
\end{align}
and
\begin{align}
 L_{1} ({\bm Q}_{12},\Omega_{12}; { \bm {\mathcal A}},\Phi)
 &=\sum_{\alpha_1\alpha_2} 
 \Big [
\Pi_{\ell; \alpha_1 \alpha_2} ({\bm Q}_{12},\Omega_{12})  
+ \Pi_{\ell; \alpha_2 \alpha_1} ({\bm Q}_{21},\Omega_{21})  
\Big ] 
{\cal A}_{\alpha_2}({\bm q}_2,\omega_2) q_{1,\alpha_1}
\nonumber \\
&- \sum_{\alpha_1} 
\Big [
\Pi_{\ell; \alpha_1 n} ({\bm Q}_{12},\Omega_{12})   \omega_1 {\cal A}_{\alpha_1}({\bm q}_2,\omega_2)
\nonumber\\ 
&+
\Pi_{\ell; \alpha_1 n} ({\bm Q}_{21},\Omega_{21})   \Phi({\bm q}_2,\omega_2) q_{1,\alpha_1}
\Big ] 
\nonumber\\
&+ 
\Big [
 \Pi_{\ell; n n} ({\bm Q}_{12},\Omega_{12})  
+
 \Pi_{\ell; n n} ({\bm Q}_{21},\Omega_{21})  
\Big ] \Phi({\bm q}_2,\omega_2) \omega_1~.
\end{align}
Because of gauge invariance, both  $\sum_{\bm q_2} L_{1}$ and $L_{2}$ must be identically zero, a fact that leads to the following relation 
for $L_{2}=0$~:
\begin{align}\label{eq:gauge_invarince_identity_j2}
 \Pi_{\ell; n n} ({\bm Q}_{12},\Omega_{12})  \omega_1\omega_2 
&= \sum_{\alpha_2}\Pi_{\ell;n \alpha_2} ({\bm Q}_{12},\Omega_{12}) \omega_1 q_{2,\alpha_2} 
\nonumber\\&-  \sum_{\alpha_1\alpha_2}\Pi_{\ell; \alpha_1 \alpha_2} ({\bm Q}_{12},\Omega_{12})  q_{1,\alpha_1} q_{2,\alpha_2}~.
\end{align}
Replacing Eq.~(\ref{eq:gauge_invarince_identity_j2}) in Eq.~(\ref{eq:j2}) we find
\begin{align}\label{eq:j2_2}
J^{(2)}_\ell ({\bm q}, \omega_\Sigma) &=
 \sum_{{\bm q}_1{\bm q}_2} 
\sum_{\alpha_1\alpha_2} 
\Pi_{\ell; \alpha_1 \alpha_2} ({\bm Q}_{12},\Omega_{12})  
 \frac{E_{\alpha_1}({\bm q}_1,\omega_1)}{i\omega_1}  \frac{E_{\alpha_2}({\bm q}_2,\omega_2)}{i \omega_2}
\nonumber\\
&+
 \sum_{{\bm q}_1 {\bm q}_2 }  \sum_{\alpha_2}  
\Big [ 
V_{\ell \alpha_2}({\bm Q}_{12},\Omega_{12})   E_{\alpha_2}({\bm q}_2,\omega_2)
\Big ]
 \frac{\Phi({\bm q}_1,\omega_1)}{i\omega_2\omega_1}~,
\end{align}
where
$V_{\ell \alpha_2}({\bm Q}_{12},\Omega_{12}) =2\sum_{\alpha_1} \Pi_{\ell; \alpha_1 \alpha_2} ({\bm Q}_{12},\Omega_{12})  
~q_{1,\alpha_1}
-\Pi_{\ell;  n \alpha_2} ({\bm Q}_{12},\Omega_{12}) ~\omega_1$.
Eq.~(\ref{eq:j2_2}) is not a fully gauge-invariant relation for the second-order current because it contains the scalar potential $\Phi$. This is because we have not yet used the condition $\sum_{\bm q_2} L_{1}=0$. However, for practical reason we instead  perform another gauge transformation and later one can show that $\sum_{\bm q_2} L_{1}=0$ is also satisfied. Accordingly, after performing a gauge transformation to the above relation we obtain 
\begin{equation}
\delta J^{(2)}_\ell ({\bm q}, \omega_\Sigma) =
 \sum_{{\bm q}_1 {\bm q}_2 } 
\left [  
\sum_{\alpha_2} 
V_{\ell \alpha_2}({\bm Q}_{12},\Omega_{12})   E_{\alpha_2}({\bm q}_2,\omega_2)
\right ] \frac{\Lambda({\bm q}_1,\omega_1)}{\omega_2  } =0~.
\end{equation}
Because of the gauge invariance property of the physical charge current, we obtain another gauge invariance identity:
\begin{equation}\label{eq:VE=0}
\sum_{\bm q_2}\sum_{\alpha_2}  V_{\ell \alpha_2}({\bm Q}_{12},\Omega_{12}) E_{\alpha_2}({\bm q}_2,\omega_2) =0~.
\end{equation}
By considering this new identity, we reach the following gauge-invariant relation of the second-order current 
\begin{equation}\label{eq:gauge_invariance_J2}
J^{(2)}_\ell({\bm q}, \omega_\Sigma) =  
 \sum_{{\bm q}_1 {\bm q}_2 } 
\sum_{\alpha_1\alpha_2} 
\Pi_{\ell; \alpha_1 \alpha_2} ({\bm Q}_{12},\Omega_{12})  
 \frac{E_{\alpha_1}({\bm q}_1,\omega_1)}{i\omega_1}  \frac{E_{\alpha_2}({\bm q}_2,\omega_2)}{i \omega_2}~.
\end{equation}
Therefore, the second-order conductivity reads as follows:
\begin{equation}
\sigma^{(2)}_{\ell\alpha_1\alpha_2}({\bm Q}_{12},\Omega_{12}) \equiv   \frac{\Pi_{\ell; \alpha_1 \alpha_2} ({\bm Q}_{12},\Omega_{12})  }{i \omega_1 i \omega_2}~.
\end{equation}
Since Eq.~(\ref{eq:VE=0}) is valid for an arbitrary electric field, we conclude that
 $V_{\ell\alpha_2}({\bm Q}_{12},\Omega_{12})=0$.
Using this relation, we can simplify Eq.~(\ref{eq:gauge_invarince_identity_j2}) as follows
\begin{align} 
\hspace{-0.37cm}
\Pi_{\ell; n n} ({\bm Q}_{12},\Omega_{12})  
&= \sum_{\alpha_1\alpha_2}\Pi_{\ell; \alpha_1 \alpha_2} ({\bm Q}_{12},\Omega_{12})  \frac{ q_{1,\alpha_1} q_{2,\alpha_2}}{\omega_1\omega_2 } 
\nonumber\\&= i^2  \sum_{\alpha_1\alpha_2} q_{1,\alpha_1} q_{2,\alpha_2} \sigma^{(2)}_{\ell\alpha_1\alpha_2}({\bm Q}_{12},\Omega_{12})~.
\end{align}
The same result can be obtained in an intuitive way similar to Eq.~(\ref{eq:N2_simpel_way}).
The above equation represents an important  Ward identity, which ensures the gauge invariance of the second-order current. Similarly to the case of the second-order density, it is easy to prove that $ \sum_{\bm q_2} L_1=0$---this result can be obtained by performing a calculation similar to that reported in~\ref{app:K1L1} for the proof of $ \sum_{\bm q_2} K_1=0$.

\begin{figure}[t]
\centering
\begin{overpic}[width=0.9\linewidth]{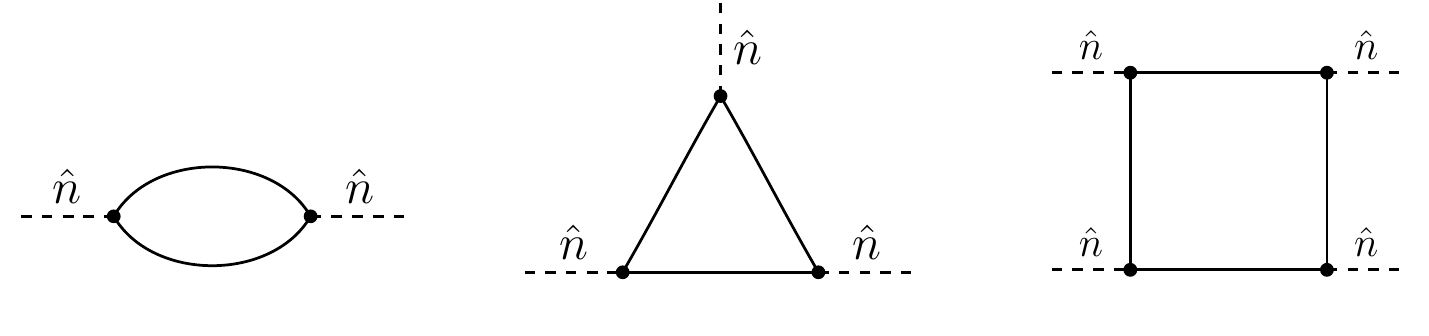}
\put(0,50){ (a)} 
\put(120,50){ (b)} 
\put(210,50){(c)}\end{overpic}
\caption{ Diagrams for nonlinear response functions in the scalar potential gauge, i.e.~$\Pi_{n;n \dots n} $. (a), (b) and (c) panels correspond to the linear, second-order, and third-order response functions, respectively. Solid lines indicate electronic propagators and solid circles stand for density vertices.
Dashed lines indicate external photon fields.}
\label{fig:diagrams_scalar}
\vspace{10mm}
\end{figure}

\begin{figure*}[h]
\centering
\begin{overpic}[width=1.1\linewidth]{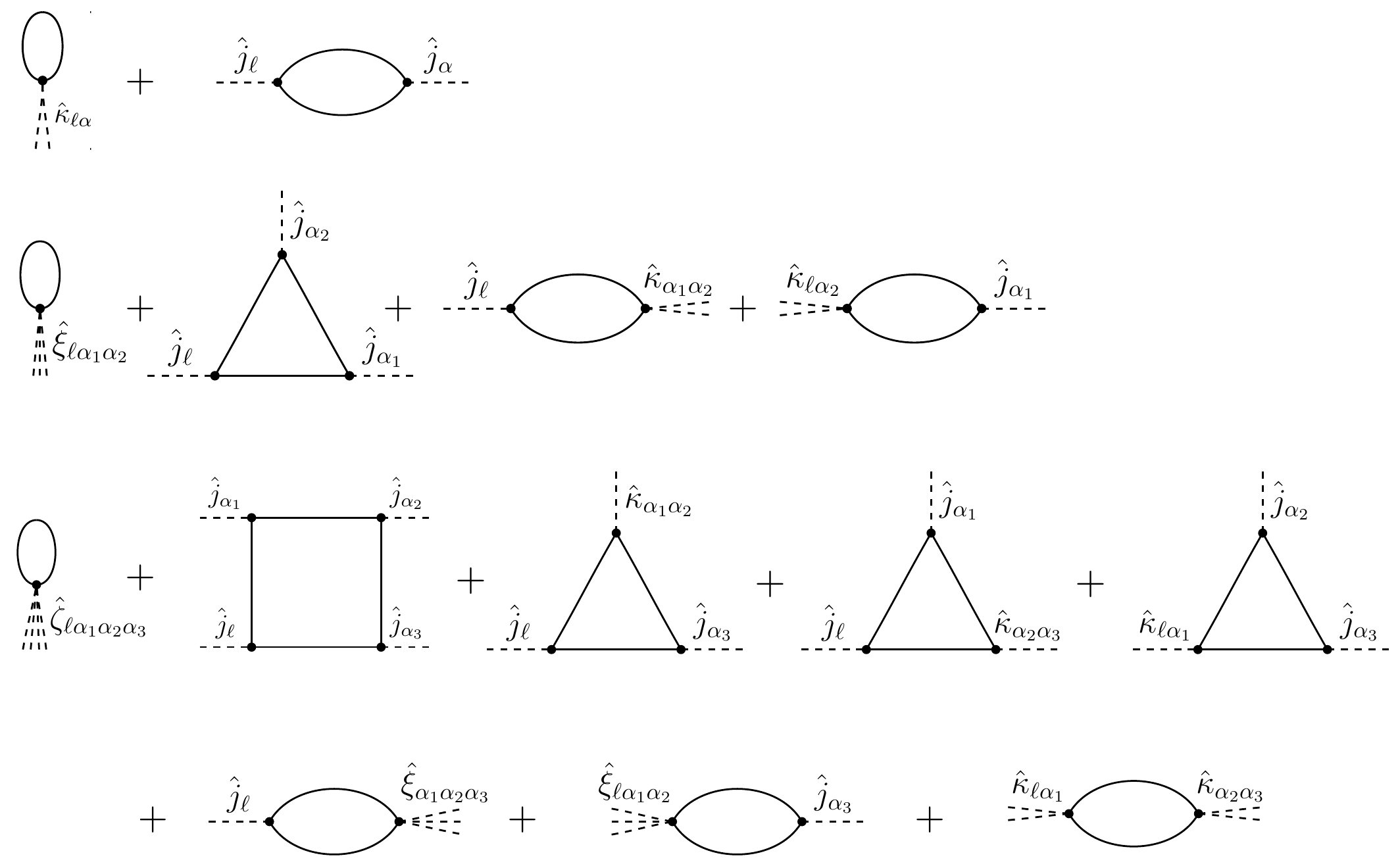}
\put(-20,195){ (a)}
\put(-20,140){ (b)}
\put(-20,70){ (c)}
\end{overpic}
\caption{ Diagrams for nonlinear response functions in vector potential gauge, i.e.~$\Pi_{\ell; \alpha_1\dots \alpha_m}$. 
Solid lines indicate electronic propagators and solid circles stand for the density and current vertices.
(a), (b) and (c) panels correspond to the linear, second-order, and third-order response function, respectively. 
Dashed lines indicate external photon fields. The overall sign of each bubble obeys the simple rule $(-1)^{p-1}$ where $p$ is the number of vertices. 
Notice that for all diagrams we must consider the intrinsic permutation symmetrization as discussed in the main text.}
\label{fig:diagrams_vector}
\end{figure*}

Finally, we visualize linear, second-order, and third-order response functions in the scalar and vector potential gauges in Figs.~\ref{fig:diagrams_scalar} and Fig.~\ref{fig:diagrams_vector}, respectively. 
As it can be seen from Fig.~\ref{fig:diagrams_scalar}, in the scalar gauge, the {\it density response functions}, $\Pi_{n;n \dots n} $, are given by only one Feynman diagram for any order of perturbation, while in the vector potential gauge, see Fig.~\ref{fig:diagrams_vector}, we have to evaluate two, four, and eight diagrams for the linear, second-order and third-order {\it current response functions}, i.e.~$\Pi_{\ell; \alpha_1\dots \alpha_m}$, respectively. 

Formal expression of these diagrams for nonlinear response functions in terms of non-interacting Green's functions are given in Ref.~\cite{Moore_prb_2019}. Using diagrammatic method, nonlinear response functions are explored in two-dimensional linear dispersive systems known as Dirac materials~\cite{Wehling_2015,rostami_prb_2017,Vandelli_prb_2019,Stepanov_acs_2020,Rostami_prr_2020,Rostami_Cappeluti_arxiv_2020,Rostami_Cappeluti_arxiv_2020_thg} and in strongly-correlated system Ref.~\cite{Katsnelson_jpcm_2010}. 
\section{Continuity relation and Ward identities}\label{sec:ward}
In the absence of external fields there is no current flow (unless the ground state carries it because of e.g. spontaneous breakdown of time-reversal symmetry).  We therefore have $\partial \langle \hat n({\bm r},t)\rangle/ \partial t =0 $ and $\langle \hat j_\alpha ({\bm r},t)\rangle=0$.
In this regard, the continuity equation can be written order-by-order in perturbation theory in the following manner:
\begin{equation}
 {\bm \nabla} \cdot {\bm J}^{(m)}({\bm r},t) = - \frac{\partial N^{(m)}({\bm r},t)}{\partial t}~.
\end{equation}
In the frequency domain we therefore have:
\begin{equation}\label{eq:continuity}
\sum_{\ell} q_{\ell}  J^{(m)}_{\ell}({\bm q},\omega_\Sigma) = \omega_\Sigma  N^{(m)}({\bm q},\omega_\Sigma)~.
\end{equation}
Using Eq.~(\ref{eq:gauge_invariance_identity_n1_j1}) in Eq.~(\ref{eq:continuity}), we have
\begin{equation}
\rchi_{n j_\ell} ({\bm q},{\bm q}',\omega) = \sum_{\alpha}  ({q_\alpha}/{\omega})
\left [  
 \rchi_{ j_{\alpha} j_{\ell} } ({\bm q},{\bm q}',\omega) 
 -
 \left \langle \hat \kappa_{\alpha\ell}({\bm q}, {\bm q}') \right \rangle
 \right]~.
 \end{equation}
Replacing the above relation in Eq.~(\ref{eq:gauge_invariance_identity_n1_j1}), we arrive at
\begin{equation}\label{linear-gauge-invariance}
\rchi_{nn}({\bm q},{\bm q}',\omega)= \sum_{\ell \alpha} \frac{q_\ell q'_\alpha}{ \omega^2 } 
 \left [
 \rchi_{ j_{\ell} j_{\alpha} } ({\bm q},{\bm q}',\omega) 
 -
 \left \langle \hat \kappa_{\ell\alpha}({\bm q}, {\bm q}') \right \rangle   
\right]~. 
\end{equation}
We therefore recover the well-known linear-response Ward identity,
\begin{equation}
\rchi_{nn}({\bm q},{\bm q}',\omega)= \frac{{\bm q}\cdot{\bm q}'}{\omega^2} \rchi_{{\cal J} {\cal J}} ({\bm q},{\bm q}',\omega)~, 
\end{equation}
where
\begin{equation}
 \rchi_{{\cal J} {\cal J}}  ({\bm q},{\bm q}',\omega) =\sum_{\ell\alpha} \frac{q_\ell q'_\alpha}{{\bm q}\cdot{\bm q'}}
  \left [
 \rchi_{ j_{\ell} j_{\alpha} } ({\bm q},{\bm q}',\omega) 
 -
 \left \langle \hat \kappa_{\ell\alpha}({\bm q}, {\bm q}') \right \rangle  
  \right]~.
\end{equation}
In a translationally-invariant system ${\bm q} = {\bm q}'$ and $\chi_{{\cal J}{\cal J}}({\bm q},{\bm q}',\omega)$ represents the {\it longitudinal} current-current response function.

Similarly to the linear-response case, we can obtain a ``second-order Ward identity" from the second-order continuity equation.  
Using Eqs.~(\ref{eq:gauge_invariance_N2}), (\ref{eq:gauge_invariance_J2}) and Eq.~(\ref{eq:continuity}), we arrive at 
\begin{align}
\sum_{{\bm q}_1 {\bm q}_2} 
\sum_{\alpha_1\alpha_2} 
&\left [ \sum_\ell q_{\ell} \Pi_{\ell; \alpha_1 \alpha_2} ({\bm Q}_{12},\Omega_{12}) -\omega_\Sigma \Pi_{n;\alpha_1\alpha_2 } ({\bm Q}_{12},\Omega_{12})  \right ]
 \nonumber\\& \times \frac{E_{\alpha_1}({\bm q}_1,\omega_1)}{i\omega_1}  \frac{E_{\alpha_2}({\bm q}_2,\omega_2)}{i \omega_2}
 =0~. 
\end{align}
We therefore conclude that 
$\Pi_{n;\alpha_1\alpha_2 } ({\bm Q}_{12},\Omega_{12}) = \sum_\ell  q_{\ell}\Pi_{\ell; \alpha_1 \alpha_2} ({\bm Q}_{12},\Omega_{12})/\omega_\Sigma $. 
Substituting this relation in Eq.~(\ref{eq:gauge_invarince_identity_n2}), we obtain the following second-order Ward identity:
\begin{equation}\label{second-gauge-invariance}
\Pi_{n;nn } ({\bm Q}_{12},\Omega_{12}) = \frac{1}{\omega_\Sigma} \sum_{\ell} q_{\ell} \sum_{\alpha_1\alpha_2} 
\frac{q_{1,\alpha_1}q_{2,\alpha_2}}{ \omega_1 \omega_2  }
 \Pi_{\ell; \alpha_1 \alpha_2} ({\bm Q}_{12},\Omega_{12})~.
\end{equation}

We can generalize the first- and second-order Ward identities, i.e.~Eqs.~(\ref{linear-gauge-invariance}) and~(\ref{second-gauge-invariance}), to the case of the $m$-th order response functions as following:
\begin{align}\label{mth_gauge_invariance}
\Pi_{n; \dots n } ({\bm q},{\bm q}_1,\dots,{\bm q}_m,\omega_1,\dots,\omega_m) &= 
\frac{(-1)^m}{\omega_\Sigma}\sum_{\ell} q_{\ell }  \sum_{ \{\alpha_i\}} \frac{q_{1,\alpha_1}\dots q_{m,\alpha_m}}{\omega_1 \dots \omega_m } 
\nonumber \\ &\times
\Pi_{\ell; \alpha_1\dots \alpha_m} ({\bm q},{\bm q}_1,\dots,{\bm q}_m,\omega_1,\dots,\omega_m)~.
\end{align}
Notice that in a translationally invariant system, we have  ${\bm q}={\bm q}_\Sigma=\sum^m_{i=1} {\bm q}_i$~. Moreover, the $m$-th order conductivity, the desirable gauge-invariant  response function, reads
\begin{align}\label{mth_conductivity}
\sigma^{(m)}_{\ell \alpha_1\dots \alpha_m}  ({\bm q},{\bm q}_1,\dots,{\bm q}_m,\omega_1,\dots,\omega_m) = 
\frac{\Pi_{\ell; \alpha_1\dots \alpha_m} ({\bm q},{\bm q}_1,\dots,{\bm q}_m,\omega_1,\dots,\omega_m)}{i\omega_1 \dots i\omega_m}~.
\end{align}

\par
It is useful to discuss the relation between nonlinear conductivities and nonlinear density response functions. 
Considering Eq.~(\ref{mth_gauge_invariance}) and Eq.~(\ref{mth_conductivity}) we find that 
\begin{align}
\Pi_{n; \dots n } ({\bm q},{\bm q}_1,\dots,{\bm q}_m,\omega_1,\dots,\omega_m) &= 
\frac{(-i)^m}{\omega_\Sigma} \sum_{\ell  \{\alpha_i\}} q_{\ell }   [q_{1,\alpha_1}\dots q_{m,\alpha_m}]
\nonumber \\ &\times
\sigma^{(m)}_{\ell \alpha_1\dots \alpha_m}  ({\bm q},{\bm q}_1,\dots,{\bm q}_m,\omega_1,\dots,\omega_m)~.
\end{align}
This relation provides a gauge-covariant prescription in order to relate the dynamical non-local conductivity to the density response function at each order of perturbation theory. 
In the optical (electric-dipole) approximation we can neglect the wave-vector dependence of the conductivity and therefore the wave-vector expansion of the density response function in the electric-dipole approximation reads as following:
\begin{align}
\Pi_{n; \dots n } ({\bm q}_1,\dots,{\bm q}_m,\omega_1,\dots,\omega_m) &\approx   
\frac{(-i)^m}{\omega_\Sigma} \sum_{\ell  \{\alpha_i\}} q_{\Sigma,\ell }   [q_{1,\alpha_1}\dots q_{m,\alpha_m}]
\nonumber\\&\times 
\sigma^{(m)}_{\ell \alpha_1\dots \alpha_m}  (\omega_1,\dots,\omega_m)~.
\end{align}
In a translationally-invariant system we have ${\bm q}={\bm q}_{\Sigma}$. The latter is therefore {\it not} an independent variable, as indicated by the notation ``$q_{\Sigma,\ell }   [q_{1,\alpha_1}\dots q_{m,\alpha_m}]$". The quantity $\sigma^{(m)}_{\ell \alpha_1\dots \alpha_m}  (\omega_1,\dots,\omega_m)$, which is a shorthand for $\sigma^{(m)}_{\ell \alpha_1\dots \alpha_m}|_{\{{\bm q}_i\to0\}}$, is the $m$-th order {\it optical} conductivity.

We can go beyond the electric-dipole approximation and consider electric-quadrupole and magnetic dipole contributions by retaining the wave-vector dependence of the conductivity up to linear order:
\begin{align}
&\Pi_{n; \dots n } ({\bm q}_1,\dots,{\bm q}_m,\omega_1,\dots,\omega_m) \approx   
\frac{(-i)^m}{\omega_\Sigma} \sum_{\ell  \{\alpha_i\}} q_{\ell }   [q_{1,\alpha_1}\dots q_{m,\alpha_m}] \times 
\nonumber \\ & \Bigg \{ \sigma^{(m)}_{\ell \alpha_1\dots \alpha_m}  (\omega_1,\dots,\omega_m)
+ 
 \sum_{i} \sum_{\alpha_i} q_{i,\alpha_i} \frac{\partial \sigma^{(m)}_{\ell \alpha_1\dots \alpha_m}  ({\bm q}_1,\dots,{\bm q}_m,\omega_1,\dots,\omega_m) }{\partial q_{i,\alpha_i}} \Big|_{\{{\bm q}_i \to 0\}}
\Bigg \}~.  
\end{align}
The above relation was used~\cite{rostami_prb_2017} in studying nonlinear plasmonic effects in graphene, which is a centro-symmetric material. Because of the latter property, higher multipole contributions are important in this material~\cite{rostami_prb_2017}. 
\section{Summary}  
In this Article, we have used the equilibrium $\hat{S}$-matrix approach in order to set up a theory that allows the calculation of nonlinear response functions in both real- and imaginary-time domains. 
We have discussed an analytical continuation procedure for the nonlinear response functions, which provides a prescription in order to obtain the retarded response function from the Matsubara one. 

A large fraction of this work was devoted to the analysis of gauge invariance in the context of the nonlinear response theory. An inhomogeneous vector potential gauge is a complete gauge to take into account contributions from all multipoles, such as electric-dipole, electric-quadruple, and magnetic-dipole contributions. As a result of gauge invariance, a set of nonlinear Ward identities are obtained.   
\section*{Acknowledgment }
This work was partially supported by the European Union's Horizon 2020 research and innovation programme under grant No.~881603 (GrapheneCore3). The work of M. I. K. is supported by European Research Council via Synergy Grant No.~854843 - FASTCORR. 
G. V. was supported by the U.S. Department of Energy (Office of Science) under Grant DE- FG02-05ER46203. H. R. acknowledges the support from the Swedish Research Council (VR 2018-04252).
%

%
\appendix
\section{Perturbation theory}\label{app:Dyson_series}
\begin{figure}[h]
\centering
\includegraphics[width=0.9\linewidth]{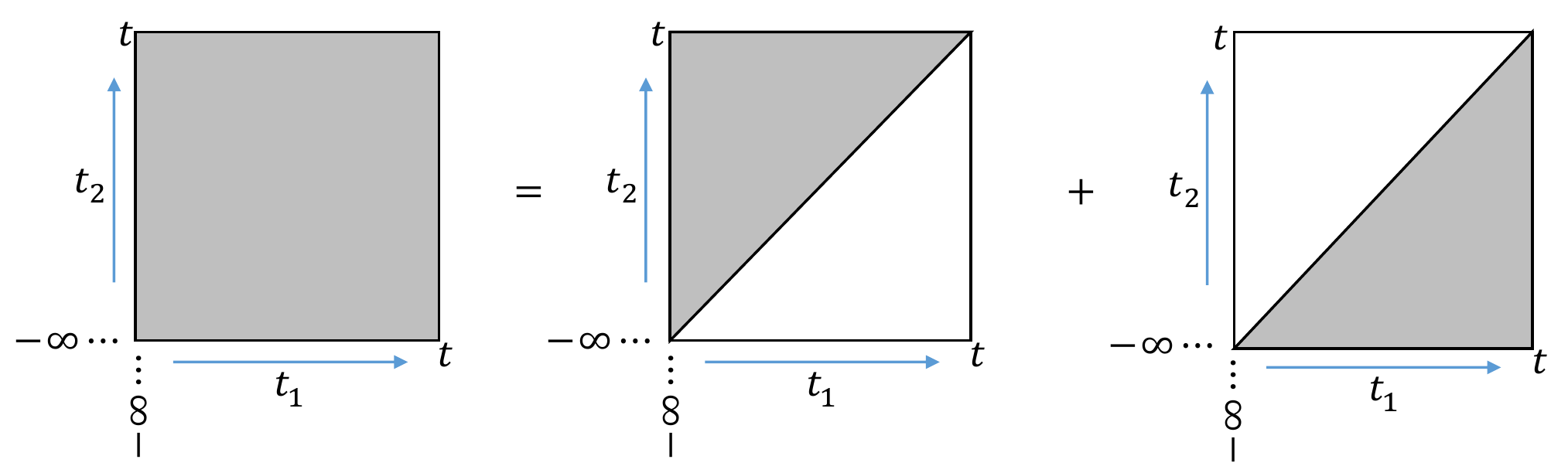}
\caption{ From the square domain into two triangle ones.}
\label{fig:time_ordering}
\end{figure}

The macroscopic value of a generic operator $\hat {\cal O}(t)$ reads as following:
\begin{align}\label{eq:O_ave_1}
{\cal O}(t) = \langle \psi(t) | \hat {\cal O}(t)|\psi(t)\rangle=\langle \psi_0 | \hat S^\dagger(t,-\infty) \hat {\cal O}(t) \hat S(t,-\infty)|\psi_0\rangle~.
\end{align}
Thanks to the Dyson expansion, we have
\begin{align}
\hat S(t,-\infty) &= 1+\sum^{\infty}_{m=1} {(-i)^m}{\cal D}^m_1 \hat V(t_1) \hat V(t_2)\dots \hat V(t_m)
\nonumber\\
\hat S^\dagger(t,-\infty) &= 1+\sum^{\infty}_{m=1} {i^m}{\cal D}^m_1 \hat V(t_m) \hat V(t_{m-1})\dots \hat V(t_1)~,
\end{align}
where we define ${\cal D}^m_1$ as follows 
\begin{align}
{\cal D}^m_1 \equiv \int^t_{-\infty} dt_1 \int^{t_1}_{-\infty} dt_2  \dots \int^{t_{m-1}}_{-\infty} dt_m~. 
\end{align}
Using Eq.~(\ref{eq:O_ave_1}), the macroscopic value of $\hat {\cal O}(t)$ is given by
\begin{align}\label{eq:O_ave_2}
  {\cal O}(t)
&=
\Big\langle 
\left\{1+i{\cal D}^1_1\hat V(t_1)+{i^2}{\cal D}^2_1 \hat V(t_2)\hat V(t_1)+{i^3}{\cal D}^3_1 \hat V(t_3)\hat V(t_2)\hat V(t_1)+\dots\right\} \hat {\cal O}(t)
\nonumber\\
&
\left\{1-i{\cal D}^1_1\hat V(t_1)+{i^2}{\cal D}^2_1 \hat V(t_1)\hat V(t_2)-{i^3}{\cal D}^3_1 \hat V(t_1)\hat V(t_2)\hat V(t_3)+\dots\right\}\Big\rangle
\nonumber\\
&=
\left\langle \hat {\cal O}(t)\right \rangle
+i{\cal D}^1_1 \left \langle\left[\hat V(t_1),\hat {\cal O}(t)\right] \right \rangle
+{i^2}\Big\langle{\cal D}^2_1 \hat V(t_2)\hat V(t_1) \hat {\cal O}(t)
\nonumber \\
&+ \hat {\cal O}(t){\cal D}^2_1 \hat V(t_1)\hat V(t_2)
-   {\cal D}^1_1\hat V(t_1)\hat {\cal O}(t)  {\cal D}^1_1\hat V(t_1)\Big\rangle
\nonumber\\
&+
{i^3}\Big\langle{\cal D}^3_1 \hat V(t_3)\hat V(t_2)\hat V(t_1) \hat {\cal O}(t)
-\hat {\cal O}(t){\cal D}^3_1 \hat V(t_1)\hat V(t_2)\hat V(t_3) 
\nonumber\\
&+ {\cal D}^1_1\hat V(t_1)  \hat {\cal O}(t) {\cal D}^2_1 \hat V(t_1)\hat V(t_2)
- {\cal D}^2_1 \hat V(t_1)\hat V(t_2) \hat {\cal O}(t) {\cal D}^1_1\hat V(t_1)
 \Big\rangle
 +\dots~.
\end{align}
In order to simplify the terms proportional to $i^2$ in the above relation, we need to rewrite ${\cal D}^1_1\hat V(t_1)\hat {\cal O}(t)  {\cal D}^1_1\hat V(t_1)$ in a proper form. 
To do so, we divide the square domain of the integral into two triangular domains (see Fig.~\ref{fig:time_ordering}). 
We therefore find
\begin{align}
{\cal D}^1_1\hat V(t_1)\hat {\cal O}(t)  {\cal D}^1_1\hat V(t_1)
&=
\int^t_{-\infty}dt_1\int^{t}_{-\infty}dt_2 \hat V(t_1)\hat {\cal O}(t)\hat V(t_2)
\nonumber\\
&=
\int^t_{-\infty}dt_1\int^{t_1}_{-\infty}dt_2 \hat V(t_1)\hat {\cal O}(t)\hat V(t_2)
\nonumber\\&+\int^t_{-\infty}dt_2\int^{t_2}_{-\infty}dt_1 \hat V(t_1)\hat {\cal O}(t)\hat V(t_2)
\nonumber\\
&=
\int^t_{-\infty}dt_1\int^{t_1}_{-\infty}dt_2 
\left\{\hat V(t_1)\hat {\cal O}(t)\hat V(t_2)+\hat V(t_2)\hat {\cal O}(t)\hat V(t_1)\right\}~.
\end{align}
Using the above identity, we reach the following compact form for the second-order perturbative terms in Eq.~(\ref{eq:O_ave_2}): 
\begin{align}\label{eq:perturb2}
&{\cal D}^2_1 \hat V(t_2)\hat V(t_1)\hat {\cal O}(t)+\hat {\cal O}(t){\cal D}^2_1 \hat V(t_1)\hat V(t_2)-{\cal D}^1_1\hat V(t_1)\hat {\cal O}(t)  
{\cal D}^1_1\hat V(t_1)
\nonumber \\
&=\int^t_{-\infty}dt_1\int^{t_1}_{-\infty}dt_2 \left[\hat V(t_2),\left[\hat V(t_1),\hat {\cal O}(t)\right]\right]~.
\end{align}
Similarly, one can find the following simplified expression for the third-order perturbation 
\begin{align}\label{eq:perturb3}
&{\cal D}^3_1 \hat V(t_3)\hat V(t_2)\hat V(t_1) {\cal O}(t)
- 
{\cal O}(t){\cal D}^3_1 \hat V(t_1)\hat V(t_2)\hat V(t_3) 
+{\cal D}^1_1\hat V(t_1)  {\cal O}(t) {\cal D}^2_1 \hat V(t_1)\hat V(t_2)
\nonumber \\
&- {\cal D}^2_1 \hat V(t_1)\hat V(t_2) {\cal O}(t) {\cal D}^1_1\hat V(t_1)
\nonumber\\
&=\int^t_{-\infty}dt_1\int^{t_1}_{-\infty}dt_2 \int^{t_2}_{-\infty}dt_3 \left[\hat V(t_3),\left[\hat V(t_2),\left[\hat V(t_1),{\cal O}(t)\right]\right]\right]~.
\end{align}
Eventually, after plugging Eqs.~(\ref{eq:perturb2}) and (\ref{eq:perturb3}) in Eq.~(\ref{eq:O_ave_2}), we arrive at the following relation for ${\cal O}(t)$:
\begin{align}
{\cal O}(t) &= \left\langle \hat {\cal O}(t)\right \rangle+i\int^t_{-\infty}dt_1\left\langle\left[\hat V(t_1), \hat {\cal O}(t)\right]\right \rangle
\nonumber\\
&+ i^2 \int^t_{-\infty}dt_1\int^{t_1}_{-\infty}dt_2 \left\langle\left[\hat V(t_2),\left[\hat V(t_1), \hat {\cal O}(t)\right]\right]\right \rangle
\nonumber\\ 
&+
i^3\int^t_{-\infty}dt_1\int^{t_1}_{-\infty}dt_2 \int^{t_2}_{-\infty}dt_3  \left\langle \left[\hat V(t_3),\left[\hat V(t_2),\left[\hat V(t_1), \hat {\cal O}(t)\right]\right]\right]\right \rangle
+\dots~.  
\end{align}
Introducing the new variables $\tau_i=t-t_i$, we have
\begin{align}
{\cal O}(t) &= \left\langle \hat {\cal O}(t)\right \rangle
+i\int^{\infty}_0d\tau_1\left\langle\left[\hat V(t-\tau_1), \hat {\cal O}(t)\right]\right \rangle
\nonumber\\
 &+ i^2 \int^{\infty}_0 d\tau_1\int^{\infty}_{\tau_1}d\tau_2 \left\langle\left[\hat V(t-\tau_2),\left[\hat V(t-\tau_1), \hat {\cal O}(t)\right]\right]\right \rangle
\nonumber\\ 
&+
i^3\int^{\infty}_0d\tau_1\int^{\infty}_{\tau_1}d\tau_2 \int^{\infty}_{\tau_2} d\tau_3  \left\langle \left[\hat V(t-\tau_3),\left[\hat V(t-\tau_2),\left[\hat V(t-\tau_1), \hat {\cal O}(t)\right]\right]\right]\right \rangle
+\dots~. 
\end{align}
Using the Heaviside function $\Theta(\tau)$, we reach 
\begin{align}\label{eq:perturbation_series}
{\cal O}(t) &= \left\langle \hat {\cal O}(t)\right \rangle + i\int^{\infty}_{-\infty}d\tau_1\Theta(\tau_1) \left\langle\left[\hat V(t-\tau_1), \hat {\cal O}(t)\right]\right \rangle
\nonumber\\ 
&+
i^2 \int^{\infty}_{-\infty} d\tau_1\int^{\infty}_{-\infty}d\tau_2~ \Theta(\tau_2-\tau_1) \Theta(\tau_1) \left\langle\left[\hat V(t-\tau_2),\left[\hat V(t-\tau_1), \hat {\cal O}(t)\right]\right]\right \rangle
\nonumber\\ 
&+
i^3\int^{\infty}_{-\infty}d\tau_1\int^{\infty}_{-\infty}d\tau_2 \int^{\infty}_{-\infty} d\tau_3 ~ \Theta(\tau_3-\tau_2)\Theta(\tau_2-\tau_1)  \Theta(\tau_1)  
\nonumber\\ &\times
\left\langle \left[\hat V(t-\tau_3),\left[\hat V(t-\tau_2),\left[\hat V(t-\tau_1), \hat {\cal O}(t)\right]\right]\right]\right \rangle
+\dots~.
\end{align}
\section{Position and real-time representation of the linear and second-order responses}\label{app:rt_o1_and_o2}
First of all, we calculate the following terms which contain different perturbative contributions. 
The average of the current operator reads
\begin{align}
\left \langle   \hat {\cal J}_{\ell}({\bm r},t;{\bm {\mathcal A}}) \right \rangle &=
 \left \langle   \hat j_{\ell}({\bm r}, t) \right \rangle+
\int d{\bm r}' \sum_{\alpha_1} \left \langle\hat \kappa_{\ell\alpha_1}({\bm r}, {\bm r}', t) \right \rangle  {\cal A}_{\alpha_1}({\bm r}',t)
\nonumber \\
&+\int d{\bm r}'\int d{\bm r}'' \sum_{\alpha_1\alpha_2} \left \langle\hat \xi_{\ell \alpha_1 \alpha_2}({\bm r}, {\bm r},{\bm r}'', t) \right \rangle  
{\cal A}_{\alpha_1}({\bm r}',t) {\cal A}_{\alpha_2}({\bm r}'',t) + \dots~.
\end{align}
 In the canonical ensemble we have $[\hat \rho, \hat{\cal H}]=0$ where $\hat\rho$ is the density matrix. Therefore, we can show that
\begin{align}
\langle \hat \kappa_{\ell \alpha_1} ({\bm r},{\bm r}',t) \rangle={\rm Tr} \left [\hat\rho e^{it \hat{\cal H}} \hat \kappa_{\ell \alpha_1} ({\bm r},{\bm r}') e^{-it \hat{\cal H}} \right ]
={\rm Tr} \left [\hat\rho \hat \kappa_{\ell \alpha_1} ({\bm r},{\bm r}') \right ] =\langle \hat \kappa_{\ell \alpha_1} ({\bm r},{\bm r}') \rangle~.
\end{align}
and similarly, $\langle \hat \xi_{\ell \alpha_1\alpha_2}({\bm r},{\bm r}',{\bm r}'',t) \rangle=\langle \hat \xi_{\ell \alpha_1\alpha_2}({\bm r},{\bm r}',{\bm r}'') \rangle$.
Eventually, we have
\begin{align}\label{eq:aveJ}
\left \langle   \hat {\cal J}_{\ell}({\bm r},t; {\bm {\mathcal A}}) \right \rangle &=
 \left \langle   \hat j_{\ell}({\bm r}, t) \right \rangle+
\int d{\bm r}' \sum_{\alpha_1} \left \langle\hat \kappa_{\ell\alpha_1}({\bm r}, {\bm r}') \right \rangle  {\cal A}_{\alpha_1}({\bm r}',t)
\nonumber \\
&+\int d{\bm r}'\int d{\bm r}'' \sum_{\alpha_1\alpha_2} \left \langle\hat \xi_{\ell \alpha_1 \alpha_2}({\bm r}, {\bm r},{\bm r}'') \right \rangle  
{\cal A}_{\alpha_1}({\bm r}',t) {\cal A}_{\alpha_2}({\bm r}'',t) +  \dots~.
\end{align}
The other required terms are the following commutators
\begin{align}\label{eq:aveVJ}
\left \langle \left[\hat V(t_1),\hat {\cal J}_{\ell}({\bm r},t;{\bm {\mathcal A}})\right] \right \rangle
&=    \int d{\bm r}_1
\Bigg \{ 
\left \langle \left [ \hat n ({\bm r}_1,t_1), \hat {\cal J}_\ell ({\bm r},t; {\bm {\mathcal A}})\right ] \right \rangle \Phi({\bm r}_1,t_1) 
\nonumber\\ 
&-
\sum_{{\alpha_1}}  \left \langle \left [ \hat {\cal J}_{\alpha_1} ({\bm r}_1,t_1; {\bm {\mathcal A}}), \hat {\cal J}_\ell ({\bm r},t; {\bm {\mathcal A}})\right ] \right \rangle  {\cal A}_{{\alpha_1}}({\bm r}_1,t_1)
\Bigg \} 
\nonumber\\
&=
 \int d{\bm r}_1
\Bigg \{ 
\left \langle \left [ \hat n ({\bm r}_1,t_1), \hat j_\ell ({\bm r},t)\right ]\right \rangle \Phi({\bm r}_1,t_1) 
\nonumber \\ 
&-
\sum_{{\alpha_1}}  \left \langle \left [ \hat j_{\alpha_1} ({\bm r}_1,t_1), \hat j_\ell ({\bm r},t)\right ]\right \rangle  {\cal A}_{{\alpha_1}}({\bm r}_1,t_1)
\nonumber\\
&+
\int d{\bm r}' \sum_{\alpha_1} \left \langle \left [ \hat n ({\bm r}_1,t_1), \hat \kappa_{\ell{\alpha_1}} ({\bm r},{\bm r}',t)\right ] \right \rangle\Phi({\bm r}_1,t_1) {\cal A}_{{\alpha_1}}({\bm r}',t)
\nonumber\\
&-
\int d{\bm r}' \sum_{{\alpha_1}{\alpha_2}} \left \langle \left [ \hat j_{\alpha_1} ({\bm r}_1,t_1), \hat \kappa_{\ell{\alpha_2}} ({\bm r},{\bm r}',t)\right ] \right \rangle {\cal A}_{{\alpha_1}}({\bm r}_1,t_1) {\cal A}_{{\alpha_2}}({\bm r}',t)
\nonumber\\
&-
\int d{\bm r}' \sum_{{\alpha_1}{\alpha_2}} \left \langle \left [ \hat \kappa_{{\alpha_1}{\alpha_2}} ({\bm r}_1,{\bm r}',t_1), \hat j_{\ell} ({\bm r},t)\right ] \right \rangle
\nonumber \\ &\times
 {\cal A}_{{\alpha_1}}({\bm r}_1,t_1) {\cal A}_{{\alpha_2}}({\bm r}',t_1)
\Bigg \}  + \dots~, 
\end{align}
in which $\hat V(t)$ is defined in Eq.~(\ref{eq:V}) as the light-matter interaction. Moreover, we calculate the following necessary term 
\begin{align}\label{eq:aveVVJ}
 \left \langle \left [\hat V(t_2),\left[ \hat V(t_1),\hat {\cal J}_{\ell}({\bm r},t; {\bm {\mathcal A}})\right] \right ] \right \rangle
&=  \int d{\bm r}_1 \int d{\bm r}_2
\Bigg \{ 
\nonumber\\ &
 \left \langle\left [\hat n({\bm r}_2,t_2), \left [ \hat n ({\bm r}_1,t_1), \hat j_\ell ({\bm r},t)\right ]\right] \right \rangle  \Phi({\bm r}_1,t_1) \Phi({\bm r}_2,t_2)
\nonumber\\ &-
\sum_{\alpha_1}  \left \langle \left[ \hat j_{\alpha_1} ({\bm r}_2,t_2), \left [ \hat n ({\bm r}_1,t_1), \hat j_\ell ({\bm r},t)\right ] \right ] \right \rangle \Phi({\bm r}_1,t_1) {\cal A}_{{\alpha_1}}({\bm r}_2,t_2)
\nonumber\\ 
&-
\sum_{{\alpha_1}}   \left \langle \left [\hat n({\bm r}_2,t_2), \left [ \hat j_{\alpha_1} ({\bm r}_1,t_1), \hat j_\ell ({\bm r},t)\right ] \right ] \right \rangle {\cal A}_{{\alpha_1}}({\bm r}_1,t_1) \Phi({\bm r}_2,t_2)
\nonumber\\ 
&+
\sum_{{\alpha_1}{\alpha_2}}   \left \langle \left [\hat j_{\alpha_2}({\bm r}_2,t_2), \left [ \hat j_{\alpha_1} ({\bm r}_1,t_1), \hat j_\ell ({\bm r},t)\right ] \right ] \right \rangle 
\nonumber\\ 
&\times {\cal A}_{{\alpha_1}}({\bm r}_1,t_1) {\cal A}_{\alpha_2} ({\bm r}_2,t_2)
\Bigg \} + \dots~. 
\end{align}
In a similar way we can write down the corresponding expressions for $\left \langle \left[\hat V(t_1),\hat n({\bm r},t)\right] \right \rangle$ and $ \left \langle \left [\hat V(t_2),\left[ \hat V(t_1),\hat n({\bm r},t)\right] \right ] \right \rangle$, which are necessary to obtain the first- and second-order densities.  
\subsection{First- and second-order currents}
By plugging Eqs.~(\ref{eq:aveJ}), (\ref{eq:aveVJ}), and~(\ref{eq:aveVVJ}) in Eq.~(\ref{eq:perturbation_series}), we obtain the following expression for the first-order current:
\begin{align}
J^{(1)}_\ell({\bm r},t) &= \int d{\bm r}' \sum_{\alpha} \left \langle \hat\kappa_{\ell{\alpha_1}}({\bm r}, {\bm r}') \right \rangle  {\cal A}_{{\alpha}}({\bm r}',t)
\nonumber\\
&+ i  \int^{\infty}_{-\infty} d\tau_1  \int d{\bm r}_1 
\Theta(\tau_1)\Bigg \{ 
\left \langle \left [ \hat n ({\bm r}_1,t-\tau_1), \hat j_\ell ({\bm r},t)\right ]\right \rangle \Phi({\bm r}_1,t-\tau_1) 
\nonumber\\
&- \sum_{{\alpha}}  \left \langle \left [ \hat j_{\alpha} ({\bm r}_1,t-\tau_1), \hat j_\ell ({\bm r},t)\right ]\right \rangle  {\cal A}_{{\alpha}}({\bm r}_1,t-\tau_1) 
\Bigg \} ~.
\end{align}
Using the linear-response definition given in Eq.~(\ref{eq:chi1}), we rewrite the previous relation as following:
\begin{align}
J^{(1)}_\ell({\bm r},t) &= \int d{\bm r}' \Bigg \{ \sum_{\alpha} 
\Big [\left \langle \hat \kappa_{\ell{\alpha}}({\bm r}, {\bm r}') \right \rangle  {\cal A}_{{\alpha}}({\bm r}',t) 
 - \int^{\infty}_{-\infty} d\tau \rchi_{j_\ell j_{\alpha}} ({\bm r},{\bm r}',\tau)   {\cal A}_{{\alpha}}({\bm r}',t-\tau) 
 \Big ]
\nonumber\\
\hspace{-2cm}
&+ \int^{\infty}_{-\infty} d\tau \rchi_{j_\ell n} ({\bm r},{\bm r}',\tau)  \Phi({\bm r}_1,t-\tau)  \Bigg \} ~.
\end{align}
In a similar way, the second-order current in position space and time domain can be written as follows
\begin{align}\label{eq:B8}
J^{(2)}_\ell({\bm r},t) &= \int d{\bm r}'  \int d{\bm r}'' 
\sum_{{\alpha_1}{\alpha_2}}
\left \langle\hat \xi_{\ell{\alpha_1}{\alpha_2}}({\bm r}, {\bm r}',{\bm r}'') \right \rangle  
{\cal A}_{{\alpha_1}}({\bm r}',t) {\cal A}_{{\alpha_2}}({\bm r}'',t)
\nonumber\\
&+ i \int^{\infty}_{-\infty} d\tau_1  \int d{\bm r}_1   \int d{\bm r}'
\Theta(\tau_1) \Bigg \{
\nonumber \\ &
\sum_{\alpha_2} \left \langle \left [ \hat n ({\bm r}_1,t-\tau_1), \hat \kappa_{\ell{\alpha_2}} ({\bm r},{\bm r}',t)\right ] \right \rangle\Phi({\bm r}_1,t-\tau_1) {\cal A}_{{\alpha_2}}({\bm r}',t)
\nonumber\\
&-
\sum_{{\alpha_1}{\alpha_2}} \left \langle \left [ \hat j_{\alpha_1} ({\bm r}_1,t-\tau_1), \hat \kappa_{\ell{\alpha_2}} ({\bm r},{\bm r}',t)\right ] \right \rangle {\cal A}_{{\alpha_1}}({\bm r}_1,t-\tau_1) {\cal A}_{{\alpha_2}}({\bm r}',t)
\nonumber\\
&-
\sum_{{\alpha_1}{\alpha_2}} \left \langle \left [ \hat \kappa_{{\alpha_1}{\alpha_2}} ({\bm r}_1,{\bm r}',t-\tau_1), \hat j_{\ell} ({\bm r},t)\right ] \right \rangle {\cal A}_{{\alpha_1}}({\bm r}_1,t-\tau_1) 
{\cal A}_{{\alpha_2}}({\bm r}',t-\tau_1) \Bigg \} 
\nonumber\\
&+
i^2  \int^{\infty}_{-\infty} d\tau_1  \int^{\infty}_{-\infty} d\tau_2 \int d{\bm r}_1 \int d{\bm r}_2
\Theta(\tau_2- \tau_1) \Theta(\tau_1) \Bigg \{ 
\nonumber \\ &
\left [\hat n({\bm r}_2,t-\tau_2), \left [ \hat n ({\bm r}_1,t-\tau_1), \hat j_\ell ({\bm r},t)\right ]\right]  \Phi({\bm r}_1,t-\tau_1) \Phi({\bm r}_2,t-\tau_2)
\nonumber\\ &-
\sum_{\alpha_1} \left[ \hat j_{\alpha_1} ({\bm r}_2,t-\tau_2), \left [ \hat n ({\bm r}_1,t-\tau_1), \hat j_\ell ({\bm r},t)\right ] \right ] \Phi({\bm r}_1,t-\tau_1) {\cal A}_{{\alpha_1}}({\bm r}_2,t-\tau_2)
\nonumber\\ 
&-
\sum_{{\alpha_1}}  \left [\hat n({\bm r}_2,t-\tau_2), \left [ \hat j_{\alpha_1} ({\bm r}_1,t-\tau_1), \hat j_\ell ({\bm r},t)\right ] \right ] {\cal A}_{{\alpha_1}}({\bm r}_1,t-\tau_1) \Phi({\bm r}_2,t-\tau_2)
\nonumber\\ 
&+
\sum_{{\alpha_1}{\alpha_2}}  \left [\hat j_{\alpha_2}({\bm r}_2,t-\tau_2), \left [ \hat j_{\alpha_1} ({\bm r}_1,t-\tau_1), \hat j_\ell ({\bm r},t)\right ] \right ] {\cal A}_{{\alpha_1}}({\bm r}_1,t-\tau_1) {\cal A}_{\alpha_2} ({\bm r}_2,t-\tau_2)
\Bigg \}~. 
\end{align}
Using Eqs.~(\ref{eq:chi1}) and (\ref{eq:chi2}), we rewrite the previous relation as follows 
\begin{align}\label{eq:B9}
J^{(2)}_\ell({\bm r},t) &=
 \int d{\bm r}_1  \int d{\bm r}_2
\Bigg \{  
\sum_{{\alpha_1}{\alpha_2}} 
\left \langle\hat \xi_{\ell{\alpha_1}{\alpha_2}}({\bm r}, {\bm r}_1,{\bm r}_2) \right \rangle  
{\cal A}_{{\alpha_1}}({\bm r}_1,t) {\cal A}_{{\alpha_2}}({\bm r}_2,t)
\nonumber\\
&+\int^{\infty}_{-\infty} d\tau
\Bigg [
\sum_{\alpha_2} 
 \rchi_{\kappa_{\ell{\alpha_2}} n} ({\bm r},{\bm r}_2,{\bm r}_1,\tau) 
{\cal A}_{{\alpha_2}}({\bm r}_2,t) \Phi({\bm r}_1,t-\tau) 
\nonumber\\
&-   \sum_{{\alpha_1}{\alpha_2}} 
 \rchi_{\kappa_{\ell{\alpha_2}}  j_{\alpha_1} } ({\bm r},{\bm r}_1,{\bm r}_2,\tau) 
 {\cal A}_{{\alpha_1}}({\bm r}_1,t-\tau) {\cal A}_{{\alpha_2}}({\bm r}_2,t)
\nonumber\\
&-   \sum_{{\alpha_1}{\alpha_2}} 
 \rchi_{ j_\ell \kappa_{{\alpha_1}{\alpha_2}} } ({\bm r},{\bm r}_1,{\bm r}_2,\tau) 
{\cal A}_{{\alpha_1}}({\bm r}_1,t-\tau)  {\cal A}_{{\alpha_2}}({\bm r}_2,t-\tau) 
\Bigg]
\nonumber\\
&+ \int^{\infty}_{-\infty} d\tau_1  \int^{\infty}_{-\infty} d\tau_2  
\Bigg [ 
\rchi_{j_\ell n n} ({\bm r},{\bm r}_1,{\bm r}_2, \tau_1,\tau_2)  
\Phi({\bm r}_1,t-\tau_1) \Phi({\bm r}_2,t-\tau_2)
\nonumber\\ &-\sum_{\alpha_1} 
\rchi_{j_\ell n j_{\alpha_1}} ({\bm r},{\bm r}_1,{\bm r}_2, \tau_1,\tau_2)  
\Phi({\bm r}_1,t-\tau_1) {\cal A}_{{\alpha_1}}({\bm r}_2,t-\tau_2)
\nonumber\\ 
&-\sum_{{\alpha_1}} 
 \rchi_{j_\ell j_{\alpha_1} n} ({\bm r},{\bm r}_1,{\bm r}_2, \tau_1,\tau_2)  
{\cal A}_{{\alpha_1}}({\bm r}_1,t-\tau_1) \Phi({\bm r}_2,t-\tau_2)
\nonumber\\ 
&+\sum_{{\alpha_1}{\alpha_2}}  
 \rchi_{j_\ell j_{\alpha_1} j_{\alpha_2}} ({\bm r},{\bm r}_1,{\bm r}_2, \tau_1,\tau_2)  
{\cal A}_{{\alpha_1}}({\bm r}_1,t-\tau_1) {\cal A}_{\alpha_2} ({\bm r}_2,t-\tau_2)
\Bigg ]
\Bigg \}~.
\end{align}
By performing the Fourier transformation with respect to time time--- Eqs.~(\ref{eq:chi1_FQ}) and (\ref{eq:chi2_FQ})---and position as explained in~\ref{app:Fourier}, we obtain the final expressions for the first- and second-order currents presented in the main text. 
\subsection{First- and second-order densities}
The first-order density is given by:
\begin{align} 
 N^{(1)}({\bm r}, t) &=
i\int^{\infty}_{-\infty}d\tau_1 \int d {\bm r}'    
\Theta(\tau_1) \left \langle\left[\hat n({\bm r}',t-\tau_1) ,\hat n({\bm r},t) \right] \right \rangle \Phi({\bm r}', t-\tau_1)
\nonumber \\
&- i \sum_{\alpha} \int^{\infty}_{-\infty}d\tau_1 \int d {\bm r}' 
\Theta(\tau_1) \left \langle\left[\hat j_\alpha({\bm r}',t-\tau_1) ,\hat n({\bm r},t) \right] \right \rangle {\cal A}_\alpha({\bm r}', t-\tau_1)~.
\end{align}
Using the linear-response definition given in Eq.~(\ref{eq:chi1}), we rewrite the previous relation as follows 
\begin{align} 
 N^{(1)}({\bm r}, t) &=
\int^{\infty}_{-\infty}d\tau_1 \int d {\bm r}'                         
\rchi_{nn}({\bm r},{\bm r}',\tau_1)
\Phi({\bm r}', t-\tau_1)
\nonumber \\
&- \sum_{\alpha} \int^{\infty}_{-\infty}d\tau_1 \int d {\bm r}' 
\rchi_{n j_\alpha} ({\bm r},{\bm r}',\tau_1)
{\cal A}_\alpha({\bm r}', t-\tau_1)~.
\end{align}
The second-order density is given by 
\begin{align} 
 N^{(2)}({\bm r}, t) &= - i \sum_{\alpha_1 \alpha_2 } \int^{\infty}_{-\infty}d \tau_1 \int d {\bm r}' \int d {\bm r}'' 
\Theta(\tau_1)\left \langle\left[\hat \kappa_{\alpha_1\alpha_2}({\bm r}',{\bm r}'',t-\tau_1) ,\hat n({\bm r},t) \right] \right \rangle 
\nonumber \\ &\times
{\cal A}_{\alpha_1}({\bm r}', t-\tau_1) {\cal A}_{\alpha_2}({\bm r}'', t-\tau_1)
\nonumber\\ 
&+ i^2 \int^{\infty}_{-\infty}d\tau_1\int^{\infty}_{-\infty}d\tau_2 \int d {\bm r}' \int d {\bm r}''  \Theta(\tau_2-\tau_1)\Theta(\tau_1)
\Bigg \{
\nonumber \\ &
\left \langle\left[\hat n({\bm r}'',t-\tau_2),\left[\hat n({\bm r}',t-\tau_1),\hat n({\bm r},t)\right]\right] \right \rangle 
\Phi({\bm r}',t-\tau_1) \Phi({\bm r}'',t-\tau_2)
\nonumber\\ 
&-  \sum_{\alpha_1} 
\left \langle\left[\hat n({\bm r}'',t-\tau_2),\left[\hat j_{\alpha_1}({\bm r}',t-\tau_1),\hat n({\bm r},t)\right]\right] \right \rangle {\cal A}_{\alpha_1}({\bm r}',t-\tau_1) \Phi({\bm r}'',t-\tau_2)
\nonumber\\ 
&- \sum_{\alpha_2} 
\left \langle\left[\hat j_{\alpha_2}({\bm r}'',t-\tau_2),\left[\hat n({\bm r}',t-\tau_1),\hat n({\bm r},t)\right]\right] \right \rangle \Phi({\bm r}',t-\tau_1) {\cal A}_{\alpha_2}({\bm r}'',t-\tau_2)
\nonumber\\ 
&+  
\sum_{\alpha_1\alpha_2}
\left \langle\left[\hat j_{\alpha_2}({\bm r}'',t-\tau_2),\left[\hat j_{\alpha_1}({\bm r}',t-\tau_1),\hat n({\bm r},t)\right]\right] \right \rangle 
\nonumber \\ &\times
{\cal A}_{\alpha_1}({\bm r}',t-\tau_1) {\cal A}_{\alpha_2}({\bm r}'',t-\tau_2)
\Bigg\}~.
\end{align}
By using the definitions given in Eqs.~(\ref{eq:chi1}) and~(\ref{eq:chi2}), we arrive at 
\begin{align} 
N^{(2)}({\bm r}, t) & = \int d {\bm r}_1 \int d {\bm r}_2 \Bigg [ -  \int^{\infty}_{-\infty}d \tau_1
\sum_{\alpha_1 \alpha_2 } \rchi_{n  \kappa_{\alpha_1 \alpha_2}}({\bm r},{\bm r}_1,\tau_1)
{\cal A}_{\alpha_1}({\bm r}_1, t-\tau_1) {\cal A}_{\alpha_2}({\bm r}_2, t-\tau_1)
\nonumber\\ 
&+ \int^{\infty}_{-\infty}d\tau_1\int^{\infty}_{-\infty}d\tau_2 
\Bigg\{ 
\rchi_{nnn}({\bm r},{\bm r}_1,{\bm r}_2, \tau_1,\tau_2)
\Phi({\bm r}_1,t-\tau_1) \Phi({\bm r}_2,t-\tau_2)
\nonumber\\ 
&-  \sum_{\alpha_1} 
\rchi_{nj_{\alpha_1} n}({\bm r},{\bm r}_1,{\bm r}_2, \tau_1,\tau_2)
{\cal A}_{\alpha_1}({\bm r}_1,t-\tau_1) \Phi({\bm r}_2,t-\tau_2)
\nonumber\\ 
&-  \sum_{\alpha_2} 
\rchi_{n n j_{\alpha_2}}({\bm r},{\bm r}_1,{\bm r}_2, \tau_1,\tau_2)
\Phi({\bm r}_1,t-\tau_1) {\cal A}_{\alpha_2}({\bm r}_2,t-\tau_2)
\nonumber\\ 
&+  \sum_{\alpha_1\alpha_2}
\rchi_{n j_{\alpha_1} j_{\alpha_2}}({\bm r},{\bm r}_1,{\bm r}_2, \tau_1,\tau_2)
{\cal A}_{\alpha_1}({\bm r}_1,t-\tau_1) {\cal A}_{\alpha_2}({\bm r}_2,t-\tau_2) 
\Bigg\}
\Bigg ]~.
\end{align}
By performing the Fourier transformation with respect to time, Eqs.~(\ref{eq:chi1_FQ}) and (\ref{eq:chi2_FQ}), and position as explained in~\ref{app:Fourier}, we obtain the final expressions for the first- and second-order densities presented in the main text. 
\subsection{Fourier transformation}\label{app:Fourier}
In order to go from the position space ${\bm r}$  to the wave-vector space ${\bm q}$, we perform a Fourier transformation.
For a general expression of the type
\begin{align}
{\cal G}({\bm r}) = \sum_{\{{\bm r}_j\}} {\cal K}({\bm r},{\bm r}_1,\dots {\bm r}_m) {\cal F}_1({\bm r}_1) \dots {\cal F}_m({\bm r}_m)~,
\end{align}
the Fourier-transformed counterpart reads as following: 
\begin{align}
{\cal G}({\bm q}) = \sum_{\{{\bm q}_j\}} {\cal K}({\bm q},{\bm q}_1,\dots {\bm q}_m) {\cal F}_1({\bm q}_1) \dots {\cal F}_m({\bm q}_m)~,
\end{align}
where 
\begin{align}
{\cal G}({\bm q}) &=\sum_{\bm r} {\cal G}({\bm r}) e^{-i {\bm q} \cdot {\bm r}}~,~~~
{\cal F}_j({\bm q}_j) =\sum_{{\bm r}_j} {\cal F}_j({\bm q}_j) e^{-i{\bm q}_j \cdot {\bm r}_j}~,
\nonumber \\
{\cal K}({\bm q},{\bm q}_1,\dots {\bm q}_m) &= \sum_{{\bm r},\{{\bm r}_j\}} {\cal K}({\bm r},{\bm r}_1,\dots {\bm r}_m)  \exp \Big ({-i {\bm q} \cdot {\bm r}+i \sum^m_j 
{\bm q}_j \cdot {\bm r}_j } \Big )~.
\end{align}
We notice that for the case of translationally-invariant system we have ${\bm q}={\bm q}_\Sigma=\sum^m_j {\bm q}_j$~.
\section{Diamagnetic contribution to the linear conductivity: Proof for Eq.~(\ref{eq:diamagnetic})}\label{app:linear_diamagnetic}
In the scalar potential gauge, we have the following relation for the first-order current:
\begin{align}
J^{(1)}_\ell ({\bm r},t) = i  \int^t_{-\infty} dt_1  \int d{\bm r}_1 \left \langle \left [ \hat n ({\bm r}_1,t_1), \hat j_\ell ({\bm r},t)\right ]\right \rangle \Phi({\bm r}_1,t_1)~.
\end{align}
By using the cyclic property of the trace, we can prove that:
\begin{align}
\left \langle\left[\hat n({\bm r}_1,t_1), \hat j_{\ell}({\bm r},t)\right] \right \rangle 
&={\rm Tr}\left [  \hat\rho  \left[\hat n({\bm r}_1,t_1), \hat j_{\ell}({\bm r},t)\right] \right ]
={\rm Tr}\left [ \left [ \hat\rho, \hat n({\bm r}_1,t_1)\right] \hat j_{\ell}({\bm r},t)  \right ]
\nonumber\\
&=\int^\beta_0 d\lambda\left \langle \left [\hat n({\bm r}_1,t_1-i\lambda ) , \hat {\cal H} \right ] \hat j_{\ell}({\bm r},t) \right \rangle ~.
\end{align}
Notice that in the last equality of the above relation, we have used the Kubo identity~\cite{Louie_and_Cohen,Clark_and_Derrick}:
\begin{align}
 \left [\hat\rho, \hat {\cal O}(t) \right ]=  \hat\rho \int^\beta_0 d\lambda  \left [\hat{\cal O}(t-i\lambda ) , \hat{\cal H} \right ]~,
\end{align}
where
\begin{align}
\hat{\cal O}(t-i\lambda)= e^{\lambda \hat{\cal H}} \hat{\cal O}(t)  e^{- \lambda \hat{\cal H}}~.
\end{align}
By performing a straightforward calculation one can show that
\begin{align}
\left \langle\left[\hat n({\bm r}_1,t_1), \hat j_{\ell}({\bm r},t)\right] \right \rangle
= \int^\beta_0 d\lambda \left \langle \left [\hat n({\bm r}_1,0) ,\hat{\cal H} \right ] \hat j_{\ell}({\bm r},t-t_1+ i\lambda ) \right \rangle~.
\end{align}
According to the continuity equation, we have
\begin{align}
\frac{d \hat n({\bm r},0)}{dt}= -i [ \hat n({\bm r},0),\hat{\cal H}]=-{\bm \nabla}\cdot \hat {\bm j}({\bm r},0)~.
\end{align}
It is important to notice that the {\it transverse} component of the current operator does not contribute enter into the continuity relation. We can therefore proceed as follows: 
\begin{align}
\left\langle\left[\hat n({\bm r}_1,t_1),\hat j_{\ell}({\bm r},t)\right]\right\rangle= - i \int^\beta_0 d\lambda \left \langle {\bm \nabla}_{1}\cdot \hat {\bm j}({\bm r}_1,0)\hat j_{\ell}({\bm r},t-t_1+ i\lambda ) \right \rangle~. 
\end{align}
Elementary vector analysis allows as to do the following manipulations:
\begin{align}
\int d{\bm r}_1 {\bm \nabla}_{1}\cdot \hat {\bm j}({\bm r}_1,0) \Phi({\bm r}_1,t_1) 
&= 
 \int d{\bm r}_1 {\bm \nabla}_{1}\cdot \left [  \hat {\bm j}({\bm r}_1,0) \Phi({\bm r}_1,t_1) \right ]
\nonumber\\& -\int d{\bm r}_1 \hat {\bm j}({\bm r}_1,0) \cdot {\bm \nabla}_{1}\Phi({\bm r}_1,t_1)~. 
\end{align}
In deriving the previous result, we have assumed that the external field vanishes at infinity and dropped the boundary term. We therefore reach
\begin{align}
\int d{\bm r}_1 {\bm \nabla}_{1}\cdot \hat {\bm j}({\bm r}_1,0) {\Phi} ({\bm r}_1,t_1) = 
\int d{\bm r}_1 {\bm j}({\bm r}_1,0) \cdot {\bm E}({\bm r}_1,t_1)~. 
\end{align}
Note that $ {\bm j}({\bm r}_1,0)$ is the longitudinal component of the current operator. By keeping in mind that $\hat j_\alpha(\bm r_1,0)$ is a longitudinal current component, we obtain
\begin{align}
J^{(1)}_{\ell}({\bm r},t) = 
\sum_{\alpha}  \int^t_{-\infty}dt_1 \int d{\bm r}_1 \int^\beta_0 d\lambda \left \langle \hat j_{\alpha}({\bm r}_1,0) \hat j_{\ell}({\bm r},t-t_1+i\lambda) \right \rangle  E_{\alpha}({\bm r}_1,t_1)~.
\end{align}
For two arbitrary $\hat A$ and $\hat B$ operators one can prove the following identity
\begin{align}
\langle \hat A(0) \hat B(t) \rangle 
&=\frac{1}{\cal Z}{\rm Tr} \left [ e^{-\beta \hat{\cal H}} \hat A(0) \hat B(t)  \right ] =
{\rm Tr} \left [ \hat \rho  e^{\beta \hat{\cal H}} \hat B(t)  e^{-\beta \hat{\cal H}} \hat A(0)  \right ] 
\nonumber\\
& =\langle \hat B(t-i\beta) \hat A(0) \rangle~.
\end{align}
We use the above identity in the following straightforward calculation
\begin{align}
&i\int^\beta_0 d\lambda\left \langle \hat j_{\alpha}({\bm r}_1,0) \hat j_{\ell}({\bm r},t'+i\lambda) \right\rangle 
=\int^{t'+i\beta}_{t'} d\tau \left \langle \hat j_{\alpha}({\bm r}_1,0) \hat j_{\ell}({\bm r}, \tau ) \right\rangle 
\nonumber \\
&=\int^{\infty}_{t'} d\tau  \left\langle \hat j_{\alpha}({\bm r}_1,0) \hat j_{\ell}({\bm r}, \tau )\right \rangle 
- \int^{\infty}_{t'+i\beta} d\tau \left \langle \hat j_{\alpha}({\bm r}_1,0) \hat j_{\ell}({\bm r}, \tau )\right \rangle 
\nonumber \\
&=
\int^{\infty}_{t'} d\tau  \left\langle \hat j_{\alpha}({\bm r}_1,0) \hat j_{\ell}({\bm r}, \tau )\right \rangle 
- \int^{\infty}_{t'} d\tau  \left\langle \hat j_{\ell}({\bm r}, \tau) \hat j_{\alpha}({\bm r}_1,0) \right\rangle 
\nonumber \\
&=
\int^{\infty}_{t'} d\tau\left \langle \left [ \hat j_{\alpha}({\bm r}_1,0) , \hat j_{\ell}({\bm r}, \tau ) \right] \right \rangle~.
\end{align}
Eventually, for $t'=t-t_1$, we can reach the following relation for the first-order current response
\begin{align}
J^{(1)}_{\ell}({\bm r},t) = 
-i\sum_{\alpha}  \int^t_{-\infty}dt_1 \int d{\bm r}_1\int^{\infty}_{t-t_1} d\tau \left \langle \left [ \hat j_{\alpha}({\bm r}_1,0) , \hat j_{\ell}({\bm r}, \tau ) \right] \right \rangle  E_{\alpha}({\bm r}_1,t_1)~. 
\end{align}
Introducing $\tau_2=\tau$ and $\tau_1=t-t_1$, we can rewrite the previous equation as 
\begin{align}
J^{(1)}_{\ell}({\bm r},t) = 
-i\sum_{\alpha}  \int^{\infty}_0 d\tau_1 \int^{\infty}_{\tau_1} d\tau_2 \int d{\bm r}_1 \left \langle \left [ \hat j_{\alpha}({\bm r}_1,0) ,\hat  j_{\ell}({\bm r}, \tau_2 ) \right] \right \rangle  E_{\alpha}({\bm r}_1,t-\tau_1) ~.
\end{align}
Using the following Fourier transformation 
\begin{align}
E_{\alpha}({\bm r}_1,t-\tau_1) = \int d\omega E_{\alpha}({\bm r}_1,\omega )  e^{-i \omega(t-\tau_1)}~,
\end{align}
we arrive at
\begin{align}
J^{(1)}_{\ell}({\bm r}, \omega ) = 
-i\sum_{\alpha}  \int^{\infty}_0 d\tau_1 e^{i \omega\tau_1} \int^{\infty}_{\tau_1}  d\tau_2 \int d{\bm r}_1 \left \langle \left [ \hat  j_{\alpha}({\bm r}_1,0) ,\hat j_{\ell}({\bm r}, \tau_2 ) \right] \right \rangle  E_{\alpha}({\bm r}_1,\omega)~. 
\end{align}
After performing an integration by parts, we arrive at
\begin{align}
J^{(1)}_{\ell}({\bm r}, \omega ) &=
-i\sum_{\alpha}   \Bigg \{ 
 \frac{e^{i \omega \tau_1}}{i \omega  } \int^{\infty}_{\tau_1}  d\tau_2 \int d{\bm r}_1\left \langle \left [ \hat  j_{\alpha}({\bm r}_1,0) ,\hat j_{\ell}({\bm r}, \tau_2 ) \right] \right\rangle  \Big |^{\tau_1\to \infty}_{\tau_1 \to 0}
 \nonumber\\
 &+
 \int^{\infty}_{0} d\tau_1 \frac{e^{i \omega \tau_1}}{i \omega}  \int d{\bm r}_1 \left \langle \left [  \hat j_{\alpha}({\bm r}_1,0) ,\hat j_{\ell}({\bm r}, \tau_1) \right] \right \rangle 
 \Bigg \} 
 E_{\alpha}({\bm r}_1,\omega) 
 \nonumber\\
 &=
\sum_{\alpha} \Bigg \{ 
i\int^{\infty}_{0}  d\tau_2 \int d{\bm r}_1 \left \langle \left [  \hat j_{\alpha}({\bm r}_1,0) ,\hat j_{\ell}({\bm r}, \tau_2 ) \right] \right \rangle  
\nonumber\\
&- i \int^{\infty}_{0} d\tau_1 e^{i \omega \tau_1}  \int d{\bm r}_1\left \langle \left [ \hat  j_{\alpha}({\bm r}_1,0) ,\hat j_{\ell}({\bm r}, \tau_1) \right] \right \rangle 
 \Bigg \} 
 \frac{E_{\alpha}({\bm r}_1,\omega) }{i \omega }~.
\end{align}
Using the linear-response definition given in Eq.~(\ref{eq:chi1}), we find the following relation for the first-order current in the scalar potential gauge:
\begin{align}\label{eq:j1_V}
J^{(1)}_{\ell}({\bm r}, \omega ) = -\sum_{\alpha} \int d{\bm r}_1 \left [ \rchi_{j_\ell j_\alpha}({\bm r},{\bm r}_1,\omega)-\rchi_{j_\ell j_\alpha}({\bm r},{\bm r}_1,0)\right ] \frac{E_{\alpha}({\bm r}_1,\omega) }{i \omega }~.
\end{align}
Equivalently, in the wave-vector space, we have:
\begin{align}\label{eq:j1_V}
J^{(1)}_{\ell}({\bm q}, \omega ) = -\sum_{{\bm q}'}\sum_{\alpha}  \left [ \rchi_{j_\ell j_\alpha}({\bm q},{\bm q}',\omega)-\rchi_{j_\ell j_\alpha}({\bm q},{\bm q}',0)\right ] \frac{E_{\alpha}({\bm q}',\omega) }{i \omega }~.
\end{align}
Comparing Eq.~(\ref{eq:gauge_ invariant_j1}) with Eq.~(\ref{eq:j1_V}), we conclude that the following gauge-invariance identity must hold true:
\begin{align}\label{eq:gauge_invariance_j1}
\langle \hat \kappa_{\ell \alpha} ({\bm q},{\bm q}' ) \rangle =  \lim_{\omega\to0} \rchi_{j_\ell j_\alpha}({\bm q},{\bm q}',\omega)~.
\end{align}
For  a longitudinal external field ${\bm E}_{\alpha}({\bm q}',\omega)||\bm q'$ , the above relation is valid at finite $\bm q,\bm q'$. However, at finite $\bm q'$ the response to a transverse field ${\bm E}_{\alpha}({\bm q}',\omega)\perp \bm q'$ is not obtainable in the scalar potential gauge. The magnetic dipole coupling is captured only in calculations carried out in the vector potential gauge. Therefore, in the case of a transverse electric field, the above sum rule is not valid for finite $\bm q'$. We note, however, that in the local $\bm q'=0$ limit, results in the scalar and vector potential gauges are identical.
\section{Continuity relations for multi-photon current operators}\label{app:diamagnetic_continuity}

The field-dependent current operator is given by 
\begin{align}
 \hat {\cal J}_\alpha ({\bm r},t;{\bm {\mathcal A}}) 
&= \hat j_\alpha({\bm r},t) +\sum_{\beta} \int d{\bm r'} \hat \kappa_{\alpha\beta}({\bm r},{\bm r}',t) {\cal A}_\beta({\bm r}',t)
\nonumber\\
&+\sum_{\beta\gamma} \int d{\bm r}' \int d{\bm r}'' \hat \xi_{\alpha\beta\gamma}({\bm r},{\bm r}',{\bm r}'',t)   {\cal A}_{\beta}({\bm r}',t) {\cal A}_{\gamma}({\bm r}'',t)
+\dots~.
\end{align}
The charge conservation law (i.e.~the continuity constitution law), i.e.~$d \hat n /dt = i[\hat {\cal H}+\hat V,\hat n]
= - {\bm \nabla} \cdot \hat{\cal {\bm J}}$, can be written more explicitly as following:
\begin{align}\label{eq:to_be_expanded_Appendix}
 -\frac{1}{i} \left [\hat{\cal H} - \sum_{\beta} \int  \left(\int^1_0 d\lambda \hat {\cal J}_\beta({\bm r},t; \lambda {\bm {\mathcal A}})\right) {\cal A}_{\beta}({\bm r},t) d{\bm r},\hat n({\bm r},t) \right ] 
= - {\bm \nabla} \cdot \hat{\cal {\bm J}}({\bm r},t;
{\bm {\mathcal A}})~.
\end{align}
We now expand the left-hand side of the above relation in terms of the vector potential (up to quadratic order):
\begin{align}\label{eq:LHS}
&\frac{1}{i} \left [\hat{\cal H} ,\hat n({\bm r},t) \right ]
-\frac{1}{i} \left [ \sum_{\beta} \int \left(\int^1_0 d\lambda \hat {\cal J}_\beta({\bm r},t; \lambda {\bm {\mathcal A}}) \right) {\cal A}_{\beta}({\bm r},t) d{\bm r},\hat n({\bm r},t) \right ]
\nonumber\\
&= \frac{1}{i} \left [\hat{\cal H} ,\hat n({\bm r},t) \right ] -\frac{1}{i}  \sum_{\beta} \int d{\bm r}' \left [ \hat {j}_\beta({\bm r}',t) ,\hat n({\bm r},t) \right ] {\cal A}_{\beta}({\bm r}',t) 
\nonumber \\
&- \frac{1}{2i}  \sum_{\beta\gamma} \int d{\bm r}'  \int d{\bm r}'' \left [ \hat {\kappa}_{\beta\gamma}({\bm r}',{\bm r}'',t) ,\hat n({\bm r},t) \right ] 
{\cal A}_{\beta}({\bm r}',t) {\cal A}_{\gamma}({\bm r}'',t)  + {\cal O} ({\cal A}^3)~.
\nonumber \\
\end{align}
Similarly, we can expand the right-hand side of Eq.~(\ref{eq:to_be_expanded_Appendix}) as following (up to quadratic order):
\begin{align}\label{eq:RHS}
{\bm \nabla} \cdot \hat{\cal {\bm J}}({\bm r},t;
{\bm {\mathcal A}}) &=  {\bm \nabla} \cdot \hat{ {\bm j}}({\bm r},t) +
  \sum_{\alpha\beta} \int d{\bm r'} \frac{\partial}{\partial r_\alpha}\hat \kappa_{\alpha\beta}({\bm r},{\bm r}',t) {\cal A}_\beta({\bm r}',t)
  \nonumber \\
&+ 
\sum_{\alpha\beta\gamma} \int d{\bm r}' \int d{\bm r}'' \frac{\partial}{\partial r_\alpha}\hat \xi_{\alpha\beta\gamma}({\bm r},{\bm r}',{\bm r}'',t)   
{\cal A}_{\beta}({\bm r}',t) {\cal A}_{\gamma}({\bm r}'',t) +{\cal O}({\cal A}^3)~.
\nonumber \\
\end{align}
The two relations given in Eq.~(\ref{eq:LHS}) and Eq.~(\ref{eq:RHS}) must be equal for an arbitrary vector potential. Equating the  zero-th order term in the two expressions we obtain the one-photon continuity relation:
\begin{equation}\label{eq:j}
\frac{1}{i} \left [\hat{\cal H} ,\hat n({\bm r},t) \right ] =  \sum_\alpha \frac{\partial}{\partial r_\alpha} \hat{ j}_{\alpha}({\bm r},t)~.
\end{equation}
Similarly, equating the first- and second-order order terms, we obtain the two- and three-photon continuity relations, 
\begin{equation}\label{eq:kappa}
 -\frac{1}{i}  \left [ \hat {j}_\beta({\bm r}',t) ,\hat n({\bm r},t) \right ]  =\sum_{\alpha} \frac{\partial}{\partial r_\alpha} \hat \kappa_{\alpha\beta}({\bm r},{\bm r}',t)
\end{equation}
and
\begin{equation}\label{eq:xi}
 -\frac{1}{2i} \left [ \hat {\kappa}_{\beta\gamma}({\bm r}',{\bm r}'',t) ,\hat n({\bm r},t) \right ] =\sum_{\alpha}\frac{\partial}{\partial r_\alpha}\hat \xi_{\alpha\beta\gamma}({\bm r},{\bm r}',{\bm r}'',t)~.
\end{equation}
In the Schr\"odinger picture of time evolution, we have  
\begin{equation}\label{eq:j}
\left [\hat{\cal H} ,\hat n({\bm r}) \right ] = - \sum_\alpha \left(-i\frac{\partial}{\partial r_\alpha} \right) \hat{ j}_{\alpha}({\bm r})
\end{equation}
\begin{equation}\label{eq:kappa}
  \left [ \hat {j}_\beta({\bm r}') ,\hat n({\bm r}) \right ]  =\sum_{\alpha} \left(-i\frac{\partial}{\partial r_\alpha} \right) \hat \kappa_{\alpha\beta}({\bm r},{\bm r}')~,
\end{equation}
and
\begin{equation}\label{eq:xi}
\frac{1}{2}\left [ \hat {\kappa}_{\beta\gamma}({\bm r}',{\bm r}'') ,\hat n({\bm r}) \right ] =\sum_{\alpha} \left(-i\frac{\partial}{\partial r_\alpha} \right)\hat \xi_{\alpha\beta\gamma}({\bm r},{\bm r}',{\bm r}'')~.
\end{equation}
By plugging Eq.~(\ref{eq:kappa}) in Eq.~(\ref{eq:xi}) we find
\begin{equation}\label{eq:xi2}
\frac{1}{2} \left [ [\hat j_\gamma({\bm r}''), \hat n({\bm r}')],\hat n({\bm r}) \right ] =\sum_{\alpha\beta}
  \left(-i\frac{\partial}{\partial r'_\beta}\right) \left(-i\frac{\partial}{\partial r_\alpha}\right) \hat \xi_{\alpha\beta\gamma}({\bm r},{\bm r}',{\bm r}'')~.
\end{equation}
Let us just focus on the two-photon (diamagnetic) coupling which, in Fourier space, reads
\begin{equation}\label{eq:kappa_f}
[ \hat {j}_{\beta}(-{\bm q}') ,\hat n({\bm q})   ]  =\sum_{\alpha} q_\alpha 
 \hat \kappa_{\alpha\beta}({\bm q},{\bm q}')~.
\end{equation}
Note that
\begin{align}
&\hat n(\bm r) = \sum_{\bm q} \hat n({\bm q}) e^{i\bm q\cdot\bm r}~,
\\
&\hat j_\beta(\bm r') = \sum_{\bm q'} \hat j_{\beta}(-\bm q') e^{-i\bm q'\cdot\bm r}~, 
\\
&\hat \kappa_{\alpha\beta}({\bm r},{\bm r}')= \sum_{\bm q,\bm q'}\hat \kappa_{\alpha\beta}({\bm q},{\bm q}') e^{i\bm q\cdot\bm r} e^{-i\bm q'\cdot\bm r'}~.
\end{align}
The Fourier  transform  
of the density operator in first quantization reads
\begin{align}\label{eq:nq}
\hat n({\bm q})= \sum_{i} e^{-i {\bm q} \cdot \hat {\bm r}_i}~,
\end{align}
where $\sum_i$ runs over all particles. Similarly, the one-photon current operator is given by
\begin{align}\label{eq:jq}
\hat j_{\beta}(\bm q) &= \frac{1}{2} \sum_{i} \left\{ v_\beta(\hat {\bm p}_i) e^{-i {\bm q} \cdot \hat {\bm r}_i} 
+   e^{-i {\bm q} \cdot \hat {\bm r}_i} v_\beta(\hat {\bm p}_i)\right\}~,
\end{align}
 where $v_\beta({\bm p}) = \partial_{p_\beta} {\cal H}({\bm p})$ is the $\beta$-th component of the velocity operator and $\hat {\bm p}_i = - i {\bm \nabla}_i$ is the momentum operator corresponding to the $i$-th particle. Commuting current and density operators as given in Eqs.~(\ref{eq:jq}) and~(\ref{eq:nq}) we then find:
\begin{align}
[ \hat {j}_{\beta}(-{\bm q}') ,\hat n({\bm q})  ]
&=  \sum_{ij} \frac{[v_\beta(\hat {\bm p}_i),e^{-i {\bm q} \cdot \hat {\bm r}_j} ]e^{i {\bm q}' \cdot \hat {\bm r}_i}  +e^{i {\bm q}' \cdot \hat {\bm r}_i} [v_\beta(\hat {\bm p}_i),e^{-i {\bm q} \cdot \hat {\bm r}_j} ] }{2} 
\nonumber\\
&=    v_\beta(-{\bm q})
\hat n({{\bm q}-{\bm q}')}~.
\end{align}

Therefore, the particle-number conservation law leads to the following relation   
\begin{equation}\label{eq:kappa_f2}
 v_\beta(-{\bm q}) \hat n({{\bm q}-{\bm q}' }) =\sum_{\alpha} q_\alpha 
 \hat \kappa_{\alpha\beta}({\bm q},{\bm q}')~.
\end{equation}
For the case of a homogenous electron liquid, by considering $\langle\hat n({\bm q}-{\bm q}')\rangle = n \delta_{{\bm q},{\bm q}' }$ with $n$ the total particle density, we find
\begin{equation}\label{eq:kappa_f3}
 n  v_\beta(-{\bm q})   
 =\sum_{\alpha} q_\alpha 
\langle \hat \kappa_{\alpha\beta}({\bm q},{\bm q})\rangle~.
\end{equation}
For a parabolic-band model, ${\cal H}({\bm p}) ={\bm p}^2/2m$, we obtain: 
\begin{equation}\label{eq:kappa_f3}
-q_\beta\frac{n}{m} 
 =\sum_{\alpha} q_\alpha 
\langle \hat \kappa_{\alpha\beta}({\bm q},{\bm q})\rangle~,
\end{equation}
which results in the well-known diamagnetic contribution 
\begin{equation}
 \langle \hat \kappa_{\alpha\beta}({\bm q},{\bm q})\rangle = - \frac{n}{m}\delta_{\alpha\beta}~.
\end{equation}
Note that in the main text we convert particle density/current to charge density/current by including the electron charge: $\hat n\to -e\hat n$, $\hat j_\alpha \to -e \hat j_{\alpha}$, and $\hat \kappa_{\alpha\beta} \to (-e)^2 \hat \kappa_{\alpha\beta}$.  
\section{Proof for $\sum_{\bf q_2} K_1=0$}\label{app:K1L1}
In order to prove the constraint $ \sum_{\bm q_2} K_1=0$, we first simplify the expression for $K_1$. From~(\ref{eq:K1}), we have
\begin{align}
K_1&=
\sum_{{\alpha_1}{\alpha_2} } \left [\Pi_{n;{\alpha_1}{\alpha_2} } ({\bm Q}_{12},\Omega_{12}) +\Pi_{n;{\alpha_2} {\alpha_1} } ({\bm Q}_{21},\Omega_{21}) \right ]
\left [ q_{1,{\alpha_1}}  {\cal A}_{\alpha_2} ({\bm q}_2, \omega_2)   \right ]
\nonumber\\ 
&- \sum_{{\alpha_2} } \left [ \Pi_{n; n {\alpha_2}}({\bm Q}_{12},\Omega_{12})  \omega_1  {\cal A}_{\alpha_2} ({\bm q}_2, \omega_2)
 +  
 \Pi_{n; n {\alpha_2}}({\bm Q}_{21},\Omega_{21}) \Phi({\bm q}_2, \omega_2) q_{1,{\alpha_2}}
  \right ]
\nonumber\\
&+
\left [ \Pi_{n;nn}  ({\bm Q}_{12},\Omega_{12}) + \Pi_{n;nn}  ({\bm Q}_{21},\Omega_{21})  \right ] 
\left [ \omega_1  \Phi({\bm q}_2, \omega_2)   \right ]~.
\end{align}
The definitions of $\Pi_{n;{\alpha_1}{\alpha_2} } $ and $\Pi_{n;nn} $ contain the intrinsic permutation symmetry operation. We can therefore simplify the above relation as follows
\begin{align}\label{eq:linear_n_Lambda}
K_1&=2 \left \{ \sum_{{\alpha_1}{\alpha_2} }  \Pi_{n;{\alpha_1}{\alpha_2} } ({\bm Q}_{12},\Omega_{12}) \left [ q_{1,{\alpha_1}}  {\cal A}_{\alpha_2} ({\bm q}_2, \omega_2)   \right ]
+ \Pi_{n;nn}  ({\bm Q}_{12},\Omega_{12}) \left [ \omega_1  \Phi({\bm q}_2, \omega_2)   \right ] \right \} 
\nonumber\\ 
&-
\sum_{{\alpha_2} } \left [ \Pi_{n; n {\alpha_2}}({\bm Q}_{12},\Omega_{12})  \omega_1  {\cal A}_{\alpha_2} ({\bm q}_2, \omega_2)
 +  
 \Pi_{n; n {\alpha_2}}({\bm Q}_{21},\Omega_{21}) \Phi({\bm q}_2, \omega_2) q_{1,{\alpha_2}}  
  \right ]~.
\end{align}
By plugging Eq.~(\ref{eq:quad_n_Lambda}) in Eq.~(\ref{eq:linear_n_Lambda}), we arrive at 
\begin{align} 
K_1&=2 \sum_{{\alpha_1}{\alpha_2} }  \Pi_{n;{\alpha_1}{\alpha_2} } ({\bm Q}_{12},\Omega_{12})  \frac{ q_{1,{\alpha_1}}  }{i\omega_2}  E_{\alpha_2} ({\bm q}_2, \omega_2) 
+2 \sum_{\alpha_2} \Pi_{n; n {\alpha_2}} ({\bm Q}_{12},\Omega_{12}) \frac{\omega_1 q_{2,{\alpha_2}}}{\omega_2} \Phi({\bm q}_2, \omega_2) 
\nonumber\\ 
&-
\sum_{{\alpha_2} } \left [ \Pi_{n; n {\alpha_2}}({\bm Q}_{12},\Omega_{12})  \omega_1  {\cal A}_{\alpha_2} ({\bm q}_2, \omega_2)
 +  
 \Pi_{n; n {\alpha_2}}({\bm Q}_{21},\Omega_{21}) \Phi({\bm q}_2, \omega_2) q_{1,{\alpha_2}}  
  \right ]~.
\end{align}
We can collect the second and third terms in the above relation and reach
\begin{align}\label{eq:linear_n_Lambda2}
K_1&=  \sum_{\alpha_2} \left \{ 2 \sum_{{\alpha_1} }  \Pi_{n;{\alpha_1}{\alpha_2} } ({\bm Q}_{12},\Omega_{12})  q_{1,{\alpha_1}}   
-  \Pi_{n; n {\alpha_2}}({\bm Q}_{12},\Omega_{12})  \omega_1  \right \}  E_{\alpha_2} ({\bm q}_2, \omega_2) 
\nonumber\\ 
&+ i \left \{  \omega_1 \sum_{\alpha_2} \Pi_{n; n {\alpha_2}} ({\bm Q}_{12},\Omega_{12})  q_{2,{\alpha_2}}
-   \omega_2 \sum_{\alpha_2} \Pi_{n; n {\alpha_2}}({\bm Q}_{21},\Omega_{21})  q_{1,{\alpha_2}}  \right \}  \Phi({\bm q}_2, \omega_2)~.
\end{align}
By considering the definition of $U_{{\alpha_2} } ({\bm Q}_{12},\Omega_{12}) $,  we can simplify Eq.~(\ref{eq:linear_n_Lambda2}) as following:
\begin{align}
K_1 &=\sum_{{\alpha_2}}U_{{\alpha_2} } ({\bm Q}_{12},\Omega_{12})  E_{\alpha_2} ({\bm q}_2, \omega_2)  
+i  \sum_{\alpha_2} \Big [ \Pi_{n; n {\alpha_2}} ({\bm Q}_{12},\Omega_{12})  \frac{ q_{2,{\alpha_2}} }{\omega_2}
\nonumber \\
&-  \Pi_{n; n {\alpha_2}}({\bm Q}_{21},\Omega_{21})  \frac{q_{1,{\alpha_2}}}{\omega_1} \Big ] \Phi({\bm q}_2, \omega_2) \omega_1\omega_2~.
\end{align}
By using Eq.~(\ref{eq:UE=0}), we reach
\begin{align}\label{eq:K1_1}
K_1=\sum_{\alpha_2} \left [ \Pi_{n; n {\alpha_2}} ({\bm Q}_{12},\Omega_{12})  \frac{ q_{2,{\alpha_2}} }{\omega_2}
-  \Pi_{n; n {\alpha_2}}({\bm Q}_{21},\Omega_{21})  \frac{q_{1,{\alpha_2}}}{\omega_1} \right ] \Phi({\bm q}_2, \omega_2) \omega_1\omega_2~.
\end{align}
By considering Eq.~(\ref{eq:U=0}) we can prove the following relation 
\begin{align}
&\sum_{{\alpha_2}} \Pi_{n;n{\alpha_2}} ({\bm Q}_{21},\Omega_{21}) \frac{q_{1,{\alpha_2}}}{\omega_1}
=2\sum_{{\alpha_1}{\alpha_2}} \Pi_{n; {\alpha_1} {\alpha_2}} ({\bm Q}_{21},\Omega_{21})  ~\frac{ q_{2,{\alpha_1}} q_{1,{\alpha_2}}} {\omega_1\omega_2} 
\nonumber\\
&=2\sum_{{\alpha_2} {\alpha_1}} \Pi_{n;  {\alpha_2} {\alpha_1}} ({\bm Q}_{21},\Omega_{21})  ~\frac{ q_{2,{\alpha_2}} q_{1,{\alpha_1}}} {\omega_1\omega_2} ~~ \text {[Note: Relabelling: ${\alpha_1} \leftrightarrow {\alpha_2}$]}
\nonumber\\
&=2\sum_{ {\alpha_1} {\alpha_2}} \Pi_{n;  {\alpha_1} {\alpha_2} } ({\bm Q}_{12},\Omega_{12})  ~\frac{ q_{2,{\alpha_2}} q_{1,{\alpha_1}}} {\omega_1\omega_2} ~~ \text {[Note: Permutation symmetry]}
\nonumber\\
&=\sum_{{\alpha_2}} \Pi_{n; n {\alpha_2}} ({\bm Q}_{12},\Omega_{12}) \frac{q_{2,{\alpha_2}}}{\omega_2}~.
\end{align}
By plugging the above relation in Eq.~(\ref{eq:K1_1}) we get $K_1=0$ and therefore $\sum_{\bm q_2} K_1=0$. Following a very similar approach, one can prove that $\sum_{\bm q_2} L_1=0$. 

\end{document}